\documentclass[aps,prd,twocolumn,floatfix,showpacs,a4paper,nofootinbib,amsmath,amssymb]{revtex4-1}
\usepackage{graphicx} 
\usepackage{dcolumn}  
\usepackage{bm}       
\usepackage{color}
\usepackage{float}    

\newcommand{\be}{\begin{equation}}
\newcommand{\ee}{\end{equation}}
\newcommand{\ba}{\begin{eqnarray}}
\newcommand{\ea}{\end{eqnarray}}
\newcommand{\en}{\sqrt{s}}
\newcommand{\jpsi}{{\rm J}/\psi}
\newcommand{\psiprime}{\psi({\rm 2S})}
\newcommand{\ups}{\rm\Upsilon}
\newcommand{\upsi}{\rm\Upsilon(1S)}
\newcommand{\upsip}{\rm\Upsilon(2S)}
\newcommand{\upsipp}{\rm\Upsilon(3S)}
\newcommand{\upsia}{\rm\Upsilon(1S+2S+3S)}
\newcommand{\pp}{pp}
\newcommand{\ppbar}{p\overline{p}}
\newcommand{\pt}{p_{\rm T}}
\newcommand{\mt}{m_{\rm T}}

\begin{document}
\title{Using the Tsallis distribution for hadron spectra in $pp$ collisions: \\ Pions and quarkonia at $\sqrt{s}=$ 5--13000 GeV}

\author{Smbat Grigoryan}
\email{Smbat.Grigoryan@cern.ch}
\affiliation{Joint Institute for Nuclear Research, 141980 Dubna, Russia}
\affiliation{A.I.Alikhanyan National Science Laboratory (YerPhI), 0036 Yerevan, Armenia}

\date{\today}
\begin{abstract}
A thermal model, based on the Tsallis distribution and blast-wave model, is proposed to compute hadron double-differential spectra 
$d^{2}N/dp_{\rm T}dy$ in $pp$ (also high-energy $p\overline{p}$) collisions. It successfully describes the available experimental 
data on pion and quarkonia ($\phi$, ${\rm J}/\psi$, $\psi({\rm 2S})$, ${\rm\Upsilon}$ family) production at energies from 
$\sqrt{s}=$~5~GeV to the LHC ones. Simple parametrizations for the $\sqrt{s}$ dependence of the model parameters are provided
allowing predictions for the yields of these particles at new collision energies.
The model can be used also for the pion Bose-Einstein correlation studies.
\end{abstract}

\pacs{24.10.Pa, 13.85.Ni, 25.75.-q}

\maketitle

\section{\label{sec1}Introduction}

Theoretical description of the hadron transverse momentum ($\pt$) and rapidity ($y$) spectra produced in proton-proton ($\pp$), 
proton-nucleus ($pA$) and nucleus-nucleus ($AA$) collisions is one of the important tasks of high-energy physics. Since its 
realization in the QCD is still not fully satisfactory (e.g., due to the parton hadronization complicated processes, 
especially at low $\pt$), 
alternative phenomenological methods are also in use. For instance, the thermal models of the stationary fireball (hadronic gas) 
with conventional Boltzmann-Gibbs distribution (BGD) are widely used to explain the hadronic abundances and $\pt$-spectra
at low $\pt$ (see, e.g.,~\cite{BRS, Becat, Ths, Shr}). At high $\pt$ the exponential BGD is not adequate since 
the spectra have a power-law form.
The thermal models with expanding (also called flowing) fireball, like the blast-wave model (BWM)~\cite{BW1, BW2}, are included
in hadron generators~\cite{Therm, Amel, Drag}. They assume the physics scenario that the initial collision creates a 
thermalized quark-gluon fireball, which expands, cools, hadronizes and goes through the chemical 
freeze-out and finally the kinetic freeze-out, when it decays into the free-streaming hadrons.
Hadron spectra are computed usually by the Cooper-Frye formula~\cite{CF} and flow-boosted BGD.
The longitudinal flow helps us to explain the $y$-spectra (see, e.g.,~\cite{BW1, BW2, BSW}), while the radial (or transverse) flow 
flattens the $\pt$-spectra and improves the data description up to $\pt$ values of several GeV/$c$ 
(see, e.g.,~\cite{BW1, BW2, Star0, Ryb, Ghosh, Chatt, Alice1}).

Recent years the thermal models employing the Tsallis distribution (TD)~\cite{Tsallis1988} have become very popular, especially 
after the LHC operation~\cite{Bed, Beck, Wilk1, Wilk2, Biro1, Tang, Conroy, Cley1, Cley2, Azmi, Wlod1, Wong1, Wong2, Wong3, Wilk3, 
Wilk4, Urm1, Wilk5, Li, Depp1, Depp2, Depp3, Zheng, Cley3, Urm2, Star1, Phen1, Phen2, CMS1, Atlas1, Alice2}\footnote
{There are hundreds of papers that develop and/or use such models. We cite only some of them which include further references.}.
Its ability to describe the charged hadron $\pt$-spectra in a large $\pt$ range \mbox{0--200}~GeV/$c$
~\cite{Wong1, Wong2, Wong3} is very impressive.
TD is a generalization of the BGD. Besides the temperature $T$ and chemical potential $\mu$
it has an additional parameter $q$ and reduces to BGD in the limit $q\rightarrow 1$. TD can be considered as a result of
averaging of the temperature fluctuations in the BGD, where $q-1$ characterizes the strength of these fluctuations~\cite{Wilk1}
(for other interpretations, see~\cite{Urm1, Wilk5}).
The relation of TD with the QCD hard-scattering formulas is discussed in~\cite{Wong2, Wong3}.
Thanks to parameter $q$, TD provides a smooth transformation of the $\pt$-spectrum shape from the nearly exponential 
form at low $\pt$, similar to BGD, to the power-law form at high $\pt$, which is the usual domain of the perturbative QCD.
Most of the TD-based models consider a stationary fireball and are devoted to the fits of hadron $\pt$-spectra in different
collisions. Papers~\cite{Tang, Cley3} use a flowing fireball of the BWM and study the radial flow effect on 
the $\pt$-spectra. In~\cite{Li, Depp3}, the $y$-spectra of charged particles are also considered 
in the two-fireball models with a longitudinal flow. 
All these studies give different values for $T$ and $q$. Parameter $q$ increases slowly with the collision energy $\en$ and 
varies in the range 1--1.2, depending on the hadron and collision types. 
Some theoretical arguments give the upper limit $q=11/9$~\cite{Beck}.

In the present paper we propose a new thermal model based on the TD and BWM with a flowing fireball.
It utilizes thermodynamically consistent version of the TD~\cite{Conroy, Cley1}
and differs from similar models by a suitable choice of the BWM ingredients (see Sec.~\ref{sec2}), allowing us
to describe the shape and normalization of the hadronic $\pt$ and $y$ spectra, measured in 
$\pp$ collisions at energies from $\en = 5$~GeV to the highest LHC one of 13~TeV and in $\ppbar$ collisions at $\en > 500$~GeV.
In our model, unlike others which also use TD, the kinetic freeze-out temperature is the same for all hadron species.
Here we consider only pions and quarkonia since the pion data are the most abundant (in terms of statistics and $\en$ 
values) and the quarkonia ($\jpsi$, $\upsi$, ...) data cover large intervals of $\pt$ and $y$, which are important 
for our fits to better fix the model parameters. 
We provide simple parametrizations for the $\en$-dependence of the model parameters allowing us to predict the pion and 
quarkonia yields $d^{2}N/d\pt dy$ in $\pp$ collisions at new energies.
Other particles as well as $pA$ and $AA$ collisions will be considered elsewhere. 

The paper is organized as follows: Sec.~\ref{sec2} gives details of the model.  
In Sec.~\ref{sec3}, we discuss the model parameters and fit procedure. 
Sections~\ref{sec4} and~\ref{sec5} are devoted to the description of pion and quarkonia data, respectively.
In the last section, our concluding remarks are given.

\section{\label{sec2}Model description}
In thermal models the single-particle invariant yield is usually defined by the Cooper-Frye 
integral over the kinetic freeze-out space-time hypersurface $\Sigma_f$~\cite{CF}
\begin{equation}
E\frac{d^3 N}{d^3 p} = \frac{g}{(2\pi)^3}\,p_\nu \int_{\Sigma_f} d^3\Sigma^\nu f(X)\, ,
\label{eq:CooperFrye}
\end{equation}
where $X = (p_\nu u^\nu - \mu)/T$. Here the integrand $f$ is the freeze-out distribution of particle four-momentum 
$p=(E,\vec{p})$ and
four-coordinate $x=(t,\vec{x})$ with temperature $T$ and $x$-dependent collective flow four-velocity 
$u$, $u_\nu u^\nu = 1$, $\mu$ is the particle chemical potential and $g=2J+1$ is its spin degeneracy factor. 
Generally, $T$ and $\mu$ may also depend on $x$, but in order to keep our model as simple as possible, we assume
that they are constant on the $\Sigma_f$. 
Then, the invariant volume $V_f$, which is called the fireball effective volume of particle production and includes the 
flow effects, could be factored out due to the Lorentz invariance in the expression for the particle total 
integrated yield~\cite{AkkSin1, Soll, Heinz, Cley4, Bron, AkkSin2}
\begin{equation}
\begin{split}
& N = \frac{g}{(2\pi)^3}\,V_f \int d^3 p\, f(\frac{E-\mu}{T})\, ,\\
& V_f = \int_{\Sigma_f} d^3\Sigma^\nu\, u_\nu(x)\, .
\end{split}
\label{eq:Ntot} 
\end{equation}

We further assume, according to the BWM~\cite{BW1, BW2}, that the fireball flow and geometry are azimuthally symmetric
and boost invariant along the longitudinal ($z$) direction, as expected at high-energy 
$\pp$ (also $\ppbar$ and central $AA$) collisions. Now, instead of the Cartesian coordinates,
it is convenient to introduce the radial vector $\vec{r}=(r \cos \phi, r \sin \phi)$ and the Bjorken longitudinal 
proper time $\tau=\sqrt{t^2 - z^2}$ and space-time rapidity $\eta=\frac{1}{2} \ln \frac{t+z}{t-z}$.
Then, the flow four-velocity could be written as~\cite{BW1}
\be
u^{\nu}=\gamma_r (\cosh \eta, ~v_r \cos \phi, ~v_r \sin \phi, ~\sinh \eta)\, ,
\label{eq:u} 
\ee
where $\gamma_r = 1/\sqrt{1 - v_{r}^2}$ and $v_r$ is the radial flow velocity.
Expressing the particle four-momentum via the $y$ and $\pt$ ($\mt = \sqrt{m^2+\pt^2}$ is transverse mass),
$p^{\nu}=(\mt\cosh y, ~\pt \cos \phi_p,  ~\pt \sin \phi_p, ~\mt\sinh y)$, we get
\be
p_{\nu}u^{\nu}=\gamma_r [\mt\cosh (y-\eta) - v_r\pt \cos (\phi_p -\phi)]\, .
\label{eq:pu} 
\ee

The hypersurface $\Sigma_f$ in the BWM is defined by the condition that the freeze-out happens
at a constant value of the proper time: $\tau=\tau_f = const$. In this case the hypersurface element
four-vector has a simple form~\cite{BW2, Amel}
\be
d^3\Sigma^{\nu}=\tau_f (\cosh \eta, ~0, ~0, ~\sinh \eta)d\eta d^2r\, .
\label{eq:dsigma} 
\ee

We fix the $\Sigma_f$ geometry as follows: in the longitudinal direction, it is limited in the interval
$-\eta_{max} < \eta < \eta_{max}$, where a maximum longitudinal flow rapidity $\eta_{max}$ is required by the finite total
energy (this breaks the exact longitudinal boost invariance). In the radial direction the upper boundary of $r$ is
given by radius $R(\eta)$ that depends on $\eta$. This dependency plays a major role in our model for the proper
description of the hadron rapidity spectra. We have tried different forms for it and found that the following simple one 
(see, e.g.,~\cite{BW2})
\be
R(\eta) = R_0 \sqrt{1-\eta^2 /\eta_{max}^2}
\label{eq:R} 
\ee
is very successful. Since $R_0$ is the radius at $\eta = 0$, the fireball gets thinner with increase of $|\eta|$.

\begin{figure}[ht]
\begin{center}
\includegraphics[width=0.95\columnwidth]{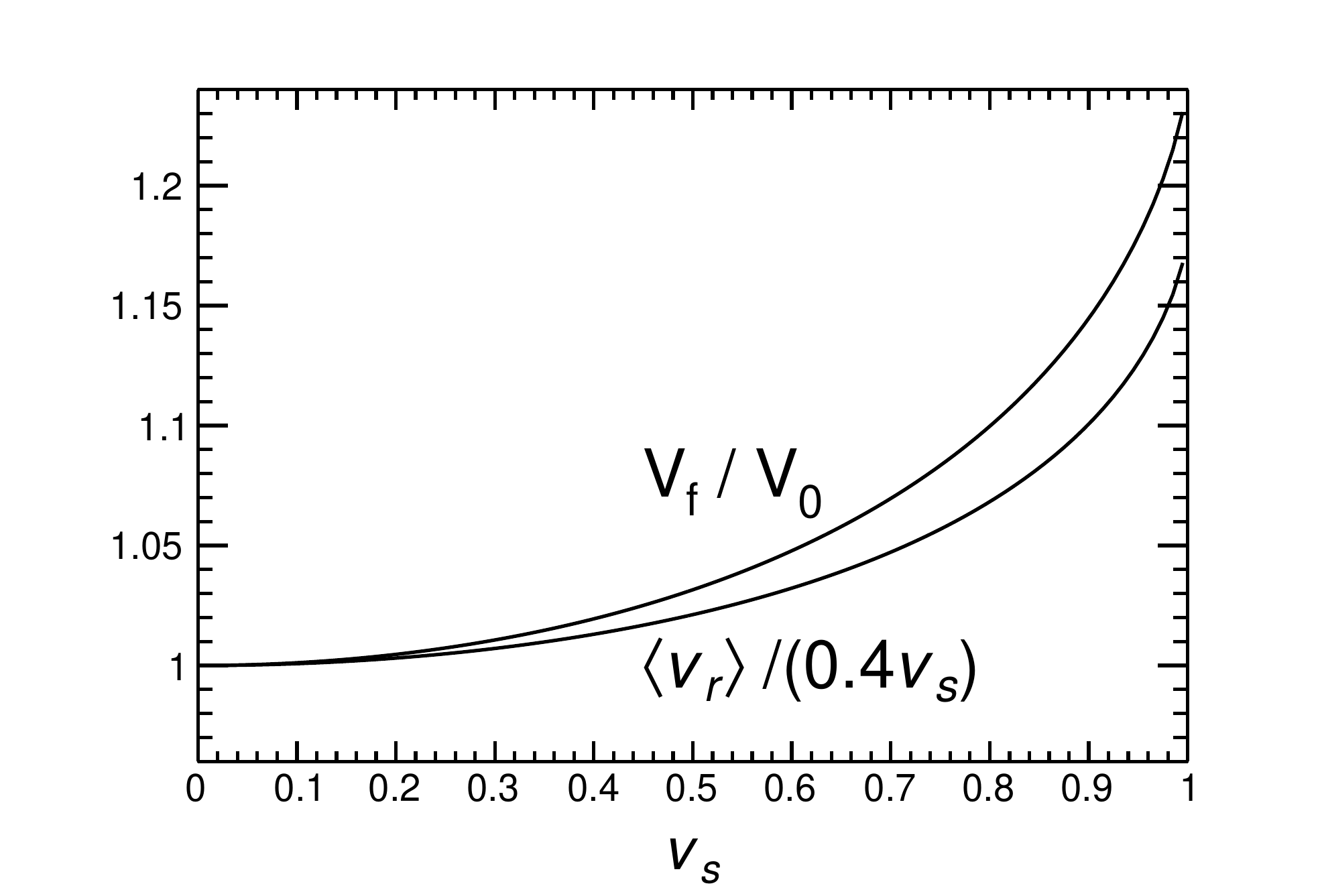}
\end{center}
\vskip -2mm
\caption{Ratios $V_f/V_0$ and $\langle v_r\rangle /(0.4 v_s)$ depending on $v_s$.}
\label{meanv_r}
\end{figure}

Now, we need to define the radial flow velocity $v_r$. Usually one assumes that it equals zero at $r=0$ 
and grows with $r$ according to a power-law dependence~\cite{BW1}. We have found that the simple quadratic dependency
\be
v_r(r) = v_s\cdot(r/R_0)^2
\label{eq:vr} 
\ee
allows us to correctly describe the hadron $\pt$ spectra. Here $v_s = v_r(R_0)$ is the surface velocity.
A useful quantity is the mean value of $v_r(r)$, which can be defined as
\be
\langle v_r\rangle = \frac{1}{V_f}\int_{\Sigma_f} d^3\Sigma^\nu\, u_\nu(x)\,v_r(r)\, .
\label{eq:meanvr1} 
\ee
According to the Eqs.~(\ref{eq:Ntot})$-$(\ref{eq:vr}) one has
\begin{eqnarray}
V_f &=& \tau_f \int_{-\eta_{max}}^{\eta_{max}} d\eta \int_0^{R(\eta)} \gamma_r r dr \int_0^{2\pi} d\phi \nonumber\\
&=& \frac{3}{2} \frac{V_0}{v_s} \int_{0}^{1} dx \arcsin[v_s (1-x^2)] \, ,
\label{eq:Vf} 
\end{eqnarray}
where $V_0 = 4/3\,\pi R_0^2 \tau_f \eta_{max}$.
Performing similar calculations with Eq.~(\ref{eq:meanvr1}) we obtain
\be
\langle v_r\rangle = \frac{3}{2} \frac{V_0}{V_f} \int_{0}^{1} dx (1-3 x^2)\arcsin[v_s (1-x^2)] \, .
\label{eq:meanvr2} 
\ee
Fig.~\ref{meanv_r} shows that the ratios $V_f/V_0$ and $\langle v_r\rangle /(0.4 v_s)$ are equal unity at $v_s = 0$
and grow with the $v_s$.

Thus, we defined the BWM ingredients of our model.
Now we specify the function $f$ in Eq.~(\ref{eq:CooperFrye}) by choosing the thermodynamically consistent TD~\cite{Conroy, Cley1}
(in contrast to the TD version, defined by Eq.~(\ref{eq:Tsallis}) with the external power index $-1$ instead of $-q$)
\be
f(X) = \left[ [1 + (q-1)X]^{\frac{1}{q-1}} - \xi \right]^{-q} ,
\label{eq:Tsallis}
\ee
where $\xi$ equals 1 or $-1$ to account for the quantum statistics of bosons or fermions, respectively.
This quantum correction matters only for pions due to their small mass. Expanding the right-hand side of 
Eq.~(\ref{eq:Tsallis}) into the binomial series and substituting it in Eq.~(\ref{eq:CooperFrye}), one gets
\begin{eqnarray}
&&E\frac{d^3 N}{d^3 p} = \frac{d^3 N}{d^2\pt dy} =
\frac{g\tau_f}{(2\pi)^3} \sum\limits_{k=0}^\infty \xi^k \binom{q-1+k}{k}\times \nonumber\\
&&\int_{-\eta_{max}}^{\eta_{max}}
d\eta \int_0^{R(\eta)} r dr \int_0^{2\pi} d\phi \frac{\mt\cosh(y-\eta)}{[1 + (q-1)X]^{\frac{q+k}{q-1}}}\, .
\label{eq:d3N} 
\end{eqnarray}
Using Eq.~(\ref{eq:pu}) and performing integrations over $\phi$ and $\phi_p$ (second integration gives $2\pi$), we obtain
\begin{eqnarray} 
&&\frac{d^2 N}{\pt d\pt dy} = g\frac{3V_0}{8\pi^2}
\sum\limits_{k=0}^\infty\xi^k\binom{q-1+k}{k}\int_{-\eta_{max}}^{\eta_{max}}\frac{d\eta}{\eta_{max}} \nonumber\\
&&\times \int_0^{R(\eta)} \frac{r dr}{R_0^2} \frac{\mt\cosh(y-\eta)\, a^{\frac{q+k}{q-1}}}
{[1 + \frac{\gamma_r \mt\cosh(y-\eta)-\mu}{T/(q-1)}]^{\frac{q+k}{q-1}}}P_{\frac{k+1}{q-1}}(a)\, ,
\label{eq:d2N} 
\end{eqnarray}
where $a=1/\sqrt{1-b^2}$\, ,
\begin{equation*}
b=\frac{\gamma_r v_r \pt}{T/(q-1) + \gamma_r \mt\cosh(y-\eta)-\mu}
\end{equation*}
and $P_\nu(a)$ is the Legendre function of the first kind~\cite{Erdelyi1}. 
Taking into account the relation ($I_0(x)$ is the modified Bessel function)
\begin{equation*}
\lim_{q \to 1}\, a^{\frac{k+1}{q-1}} P_{\frac{k+1}{q-1}}(a) = I_0(\frac{k+1}{T}\gamma_r v_r \pt)
\end{equation*}
one can easily verify that in the limit $q\rightarrow 1$ Eq.~(\ref{eq:d2N}) reproduces usual BWM formulas~\cite{BW2}
based on the BGD. Eq.~(\ref{eq:d2N}) (with Eqs.~(\ref{eq:R}) and (\ref{eq:vr})) is the main formula of our model. 
We have checked that the series in this formula is convergent if $\mu < m$ (like for similar series in the 
thermal models with BGD~\cite{Shr}). This condition is fulfilled according to Eqs.~(\ref{eq:mupar}) and (\ref{eq:mupi}).
Higher terms of the series are important only for pions (mostly for $\pi^+$ which has larger $\mu$) at low 
values of $\en,\, |y|$ and $\pt$. For example, for the case of $\en = 30.6$~GeV, $y = 0$ and $\pt \sim 0$,
considered in Sec.~\ref{sec4}, first three terms of the series give together about 97\% of the $\pi^+$ yield.
At lower energies, more terms of the series should be used for the accurate computation of pion yields.
For heavier hadrons, one can safely use $\xi=0$.

\section{\label{sec3}Parameters and fit procedure}

Here, we utilize Eq.~(\ref{eq:d2N}) for fitting the hadron spectra in $\pp$ and $\ppbar$ collisions.
We follow two aims. First is to show that our model with a possibly minimum number of parameters is
able to describe well the available data on $\pt$ and $y$ spectra for different particles and energies $\en$.
The second aim is to systematize the fit results
for different $\en$ and provide simple parametrizations for the $\en$-dependence of model parameters,
permitting predictions for the future experiments.

To fit the data given in terms of the cross section $\sigma$, we convert it to the invariant yield $N$ via the
relation $\sigma = N\sigma_{in}$, where $\sigma_{in}$ is the $\pp$ or $\ppbar$ inelastic cross section at the energy
$\en$/GeV ($\lambda$ equals 1 or $-$1, respectively, see the L2 model of Table~B1 in~\cite{sigInel})
\be
\sigma_{in} = (26.2 + 0.1717\ln^2 \frac{s}{3.521} + \frac{53.2}{s^{0.40}} - \frac{27.0}{s^{0.48}}
 - \lambda\frac{33.8}{s^{0.545}})\mbox{mb}\, .
\label{eq:siginel}
\ee
As in other applications of TD for inclusive pions, we do not calculate explicitly the feed-down 
contribution from the resonance decays,
assuming that directly produced pions and secondary ones have the same spectral shapes.
Secondary pions are expected to dominate at low $\pt$ (see, e.g.,~\cite{Star0}). 

Eq.~(\ref{eq:d2N}) has six independent parameters: $T$, $q$, $\mu$, $V_0$, $\eta_{max}$ and $v_s$. Generally, they can
depend on the $\en$ and hadron type. We assume that the kinetic freeze-out temperature T is the same for all hadron species
(the chemical freeze-out temperature may rise with the hadron mass).
Since the neutral pions and quarkonia ($\phi$, $\jpsi$, $\psiprime$, $\ups$ family) do not have conserved quantum numbers, their
chemical potential $\mu$ must equal zero in the chemical equilibrium~\cite{BRS}. We have verified that at $\en > 50$~GeV
the neutral and charged pion data can be successfully fitted with $\mu=0$, while this is not true for heavier hadrons.
The nonzero $\mu$ can be interpreted as a measure of the non-equilibrium for the given particle.
A similar fact is well known in the non-equilibrium thermal models based on BGD, where one introduces so-called 
phase-space occupancy $\gamma$, related to the chemical potential as $\mu = T\ln\gamma$~\cite{Shr}. 
To ensure the same yield for the pion three charge states at high energies, as follows from the data, we assume 
that all the model parameters, except $\mu$, are the same for these states. Moreover, we will use for them a common 
averaged mass $m_{\pi} = (2 m_{\pi^{\pm}}+m_{\pi^0})/3$.

\begin{figure}[ht]
\begin{center}
\includegraphics[width=0.95\columnwidth]{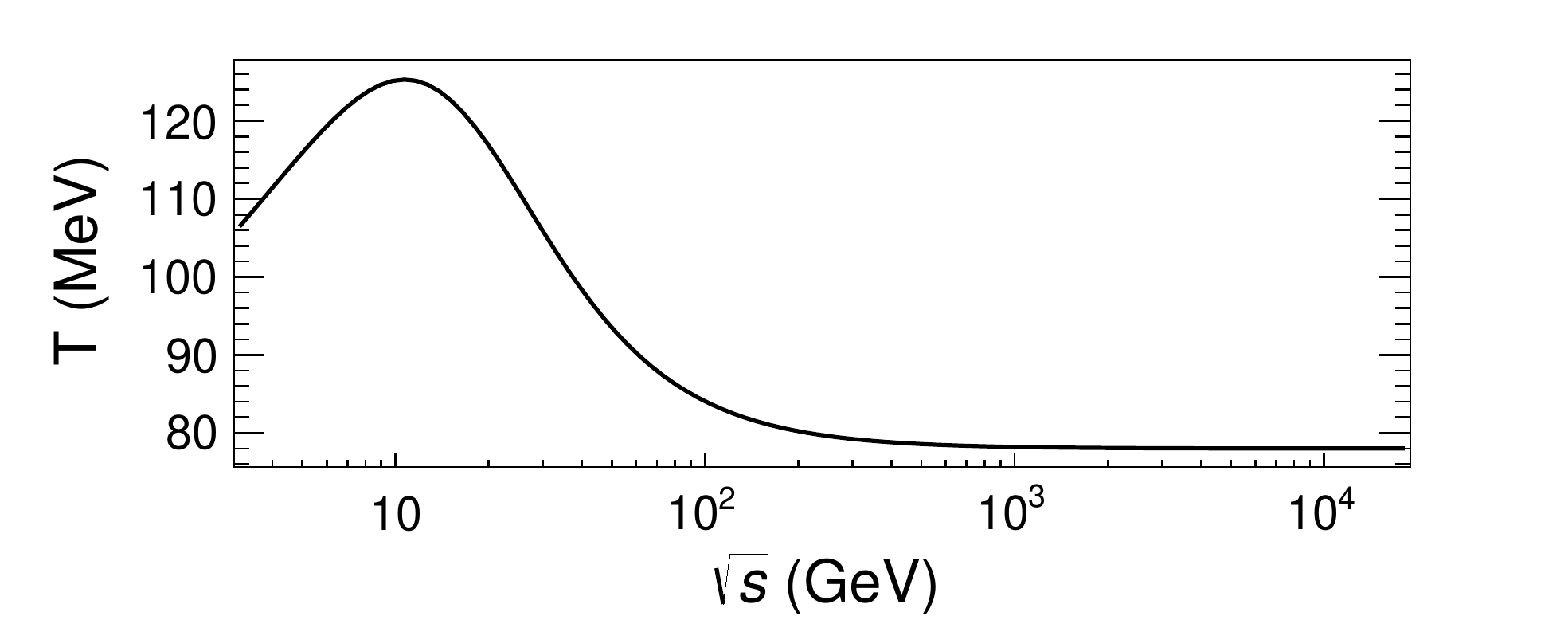}
\end{center}
\vskip -5mm
\caption{Kinetic freeze-out temperature $T$ depending on $\en$.}
\label{T_vsEn}
\end{figure}

Using the above-mentioned assumptions, we have done $\chi^2$ fits (in the ROOT framework~\cite{ROOT}) 
of the existing data on pion and quarkonia $\pt$ spectra for different values of $y$ and $\en$.
We started with the pion fits and have observed that parameter $T$ increases with energy at low energies up to about $\en = 10$~GeV.
Then, it decreases and becomes practically constant at $\en > 500$~GeV.
This behavior can be parametrized as (see Fig.~\ref{T_vsEn})
\be
T = T_{\infty}(1 + \frac{1.33\sqrt{x} - 0.21}{1+x^2})\, ,
\label{eq:T}
\ee
where $x = \en/(16$~GeV) and $T_{\infty} = 78$~MeV is the temperature at $\en \to \infty$.
Similar energy dependence was observed for the kinetic freeze-out temperature in $AA$ collisions
using thermal models with the BGD (see, e.g., Fig.~11a in~\cite{Chatt}).

We then utilized Eq.~(\ref{eq:T}) in the fits of all hadrons.
The fit results for $\eta_{max}$ and $v_s$ are parametrized as
\begin{eqnarray} 
\label{eq:eta} 
\eta_{max} &=& 0.89y_m - 0.32 - 1.18\frac{y_m}{y_b} + \frac{1.86}{y_b} - \frac{0.17}{y_m} \nonumber\\
&-& 0.025\frac{m_p}{m}\,e^{-{\en}/{e_0}}\, ,\\
v_s &=& 0.78(1 - \frac{1.31}{y_m} - \frac{0.09}{y_m}\frac{m}{m_p} - \frac{0.023}{y_b}\frac{m_p}{m})\, ,
\label{eq:etav} 
\end{eqnarray}
where $e_0 = 45$~GeV, $m$ is the mass of given hadron, $y_m = \ln(\en/m)$ its maximum rapidity, $m_p$ the proton mass and 
$y_b = \ln(\en/m_p)$ the beam rapidity in high energy $\pp$ or $\ppbar$ collisions.
In our model, the $y$-spectrum width is proportional to $\eta_{max}$ and grows logarithmically with $\en$.
Besides, the larger $m$ is, the smaller $\eta_{max}$ is and hence the narrower the $y$-spectrum is. 
Parameter $v_s$ changes in the range \mbox{0--0.78}, increases with $\en$ and decreases with increasing $m$.
Eqs.~(\ref{eq:meanvr2}) and (\ref{eq:etav}) show that radial flow velocity for pions is significant even at $\en \sim 5$~GeV
(while it vanishes at $\en \le 540$~GeV in~\cite{Tang}).
Note that the last terms in Eqs.~(\ref{eq:eta}) and (\ref{eq:etav}) are important only for pions at low energies.

The remaining fit parameters also demonstrate properties common for different hadrons.
The volume parameter can be expressed as
\be
V_0 = \widetilde{V}\, \eta_{max}\, y_b\, (T_{\infty}/T)^{2.06}\, ,
\label{eq:V0} 
\ee
where $\widetilde{V}$ is $\en$-independent but strongly decreases with increase of the hadron mass 
(see Table~\ref{tab:tab1})\footnote{
In principle, it is possible to redefine the Eq.~(\ref{eq:d2N}) parameters and obtain for heavier hadrons the same $\widetilde{V}$ 
as for pions using the following identity transformation of the TD (see also~\cite{Cley2}):
\begin{equation*}
\widetilde{V}[1+\frac{E-\mu}{T/(q-1)}]^{\frac{q}{1-q}} = \widetilde{V}_\pi[1+\frac{E-\mu'}{(T-T\delta)/(q-1)}]^{\frac{q}{1-q}}\, ,
\label{eq:scaling} 
\end{equation*}
where $\mu' = \mu - T\delta/(q-1)$ and parameter $\delta=1-(\widetilde{V}/\widetilde{V}_\pi)^{1-1/q}$ grows 
with $\widetilde{V}_\pi/\widetilde{V}$.
As mentioned in Sec.~\ref{sec1}, the $q-1$ characterizes the temperature fluctuations around the mean value $T$. 
According to~\cite{Wilk2}, the quantity $T\delta$ can be interpreted as a measure of the energy transfer, caused by these 
fluctuations, from the fireball region where the particle is produced to the surrounding regions. 
Note that $\delta \sim (q-1)$ at $q \to 1$, as expected in~\cite{Wilk2}.
}. 
Note that $V_0 \sim {\ln}^2{s}$ at high $\en$. 
The normalization constant $\widetilde{V}$ for inclusive pions, given in Table~\ref{tab:tab1}, includes the contribution 
of the resonance decays and hence is expected to be larger than the one for the directly produced pions.

\begin{table}
\caption{\label{tab:tab1}
Parameters of Eqs.~(\ref{eq:V0})$-$(\ref{eq:mupar}) for $\pi^{\pm , 0}$, $\phi$, $\jpsi$, $\psiprime$ and $\upsi$, obtained from 
the combined fits of data measured at different energies. 
The $\chi^2$ and $NDF$ correspond to the fits when all the parameters, except $\widetilde{V}$, are fixed to their central values.
Additional parameters for higher $\ups$ states and non-prompt $\jpsi$ and $\psiprime$ production are given in Sec.~\ref{sec5}.}
\begin{ruledtabular}
\begin{tabular}{cccccc}
  & $\pi$ & $\phi$ & $\jpsi$ & $\psiprime$ & $\upsi$ \\
\colrule
$\widetilde{V}$ (${\rm GeV}^{-3}$) & 
\begin{tabular}{c} 5030.9 \\ $\pm$7.2 \\ \end{tabular} &
\begin{tabular}{c} 561.2  \\ $\pm$5.2 \\ \end{tabular} &
\begin{tabular}{c} 107.4  \\ $\pm$0.4 \\ \end{tabular} & 
\begin{tabular}{c}  19.2  \\ $\pm$0.1 \\ \end{tabular} & 
\begin{tabular}{c} 0.241 \\ $\pm$0.001 \\ \end{tabular} \\
$e_1$ (GeV) & 12.5   & 12.5   & 7.8    & 7.8     & 30.0 \\
$p_1$      & 3.5     & 3.5  & 76.1    & 128.1   &  0  \\
$p_2$      & 2.3    & 3.1    &  0      &  0    &  0  \\
$p_3$      & 135.6  & 166.3 & 56.3   & 56.3   & 3.0  \\
$p_4$      & 0      & 0     & 29.2    & 29.2   & 8.6  \\
$p_5$      & 46.7  & 46.7  & 87.9    & 94.0     & $-$4.3 \\
$e_2$ (GeV) & -    & 8786   & 13900   & 35500  & 1$\cdot 10^{9}$ \\
$e_3$ (GeV) & -     & 225   & 63.1    & 63.1   & 16000  \\
$p_6$      & -     & 0.047 & 0.072   & 0.060   & 0.058  \\
$p_7$      & -     & 2.30  & 2.01    & 0.50     & 3.50 \\
\colrule
$\chi^2$   & 8716  & 235   & 5985   & 1567  & 5773 \\
$NDF$      & 2293   & 223  & 1707    & 974  & 622 \\ 
\end{tabular}
\end{ruledtabular}
\end{table}

The fitted values of $q$ grow with $\en$ and are different for different hadron species. However they vary in the very small 
interval 1--11/9, as noted in Sec.~\ref{sec1}. Therefore, it is more convenient to use the parameter $n$ instead
\be
n = q/(q-1)\, ,
\label{eq:qn} 
\ee
which controls the large-$\pt$ behavior of Eq.~(\ref{eq:d2N}). Then, $n>n_{\infty}=$11/2~\cite{Beck} and $n\to\infty$ at $q\to 1$.
the resulting fitted values of $n$ can be parametrized for different hadron species by the formula, valid at $x=e_1/\en < 1$,
\be
n = \frac{n_{\infty}}{1+p_1x} + \frac{p_3/\ln x}{\ln x - p_2} + p_4 x^{0.37} - p_5 x\, ,
\label{eq:npar} 
\ee
where parameters $e_1$ and $p_1$--$p_5$ are listed in Table~\ref{tab:tab1} and $e_1$
\begin{figure}[H]
\begin{center}
\includegraphics[width=0.95\columnwidth]{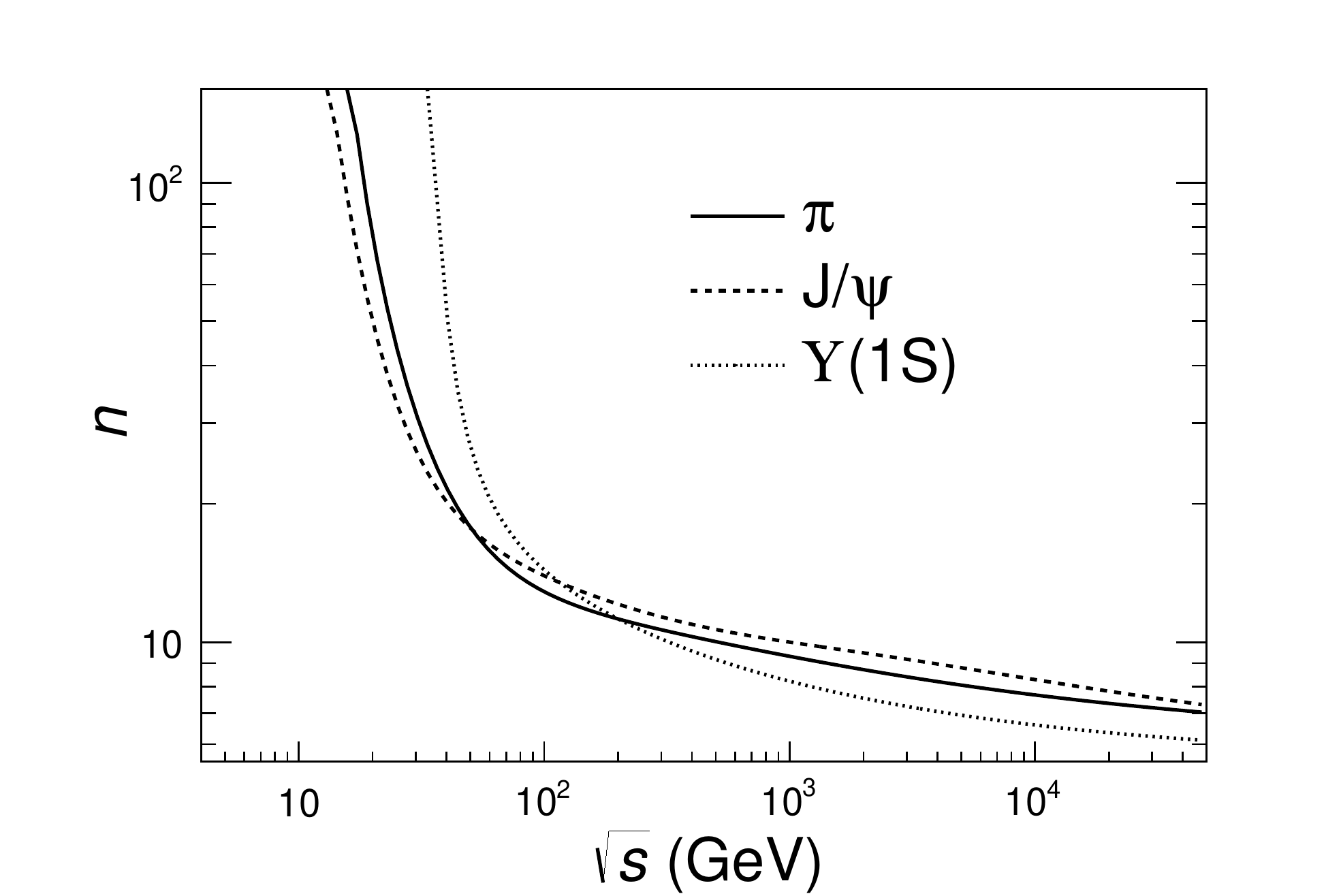}
\end{center}
\vskip -5mm
\caption{Parameter $n$ for $\pi$, $\jpsi$ and $\upsi$ depending on $\en$.}
\label{n_vsEn}
\end{figure}

\noindent
is the energy when $n$ becomes infinity. So, at $\en \le e_1$, the TD reduces to BGD.
The limiting value $n_{\infty}$ provides that Eq.~(\ref{eq:d2N}) at $\en \to \infty$ has the 
same large-$\pt$ behavior as the jet production in the lowest-order perturbative QCD~\cite{Wong3}.
Fig.~\ref{n_vsEn} shows the energy dependence of $n$ for $\pi$, $\jpsi$ and $\upsi$. The corresponding curves for
$\phi$, $\psiprime$ and higher $\ups$ states are similar to the ones for $\pi$, $\jpsi$ and $\upsi$, respectively.

The obtained values of $\mu$ are always smaller $m$. They are proportional to $m$ and vanish with increasing $\en$. 
We parametrize $\mu$ for different quarkonium species by the formula, valid at $\en \le e_2$,
\be
\mu = p_6(\ln{\frac{e_2}{\en}} - p_7\frac{1-\en/e_2}{1+\en/e_3})m\, ,
\label{eq:mupar} 
\ee
where parameters $e_2$, $e_3$, $p_6$ and $p_7$ are given in Table~\ref{tab:tab1}
and $\mu=0$ at $\en > e_2$ ($e_2$ is larger for heavier particles).
For pions we use
\be
\mu_{\pi} = \frac{2 \,\rm GeV}{\en}\,(1+0.6\,Q_{\pi})\,e^{-{\en}/{e_0}}\, m_{\pi}\, ,
\label{eq:mupi} 
\ee
where $Q_{\pi}=0$ for $\ppbar$ collisions while for $\pp$ collisions $Q_{\pi}$ equals the pion charge, to 
account for the difference of $\pi^+$, $\pi^0$ and $\pi^-$ yields in low energy $\pp$ collisions, 
related to the charge-conservation effects.
$\mu_{\pi}$ vanishes with increasing energy, in agreement with the fact that
these yields almost coincide at $\en \ge 62.4$~GeV~\cite{Phen1,Phen2}.

In Sections ~\ref{sec4} and \ref{sec5} we will discuss in more detail the results of combined fits of pion and quarkonia 
data using Eqs.~(\ref{eq:d2N})$-$(\ref{eq:mupi}).
The parameter values as well as the $\chi^2$ and $NDF$ of the fits for each hadron type are given in Table~\ref{tab:tab1}.
Additional parameters for $\jpsi$ and $\psiprime$, produced via bottom hadron decays, and for higher $\ups$ states
will be considered in Sec.~\ref{sec5}.
Note that rather large ratios $\chi^2 / NDF$ are 
due to the large amount of data included in the fits, which use $\en$-dependent parametrizations 
for the model parameters. 
Since the quality and normalization of different measurements for given hadron do not always agree well with each other,
the combined fit gives larger $\chi^2 / NDF$ than the individual fits for each measurement.
To get not-too-large $\chi^2 / NDF$, we have excluded some data samples from the combined fits.

\section{\label{sec4}neutral and charged pions}
Here, we present the results of the combined fit of
\begin{widetext}

\begin{figure}[ht]
\begin{center}
\resizebox{0.53\columnwidth}{!}
{\includegraphics{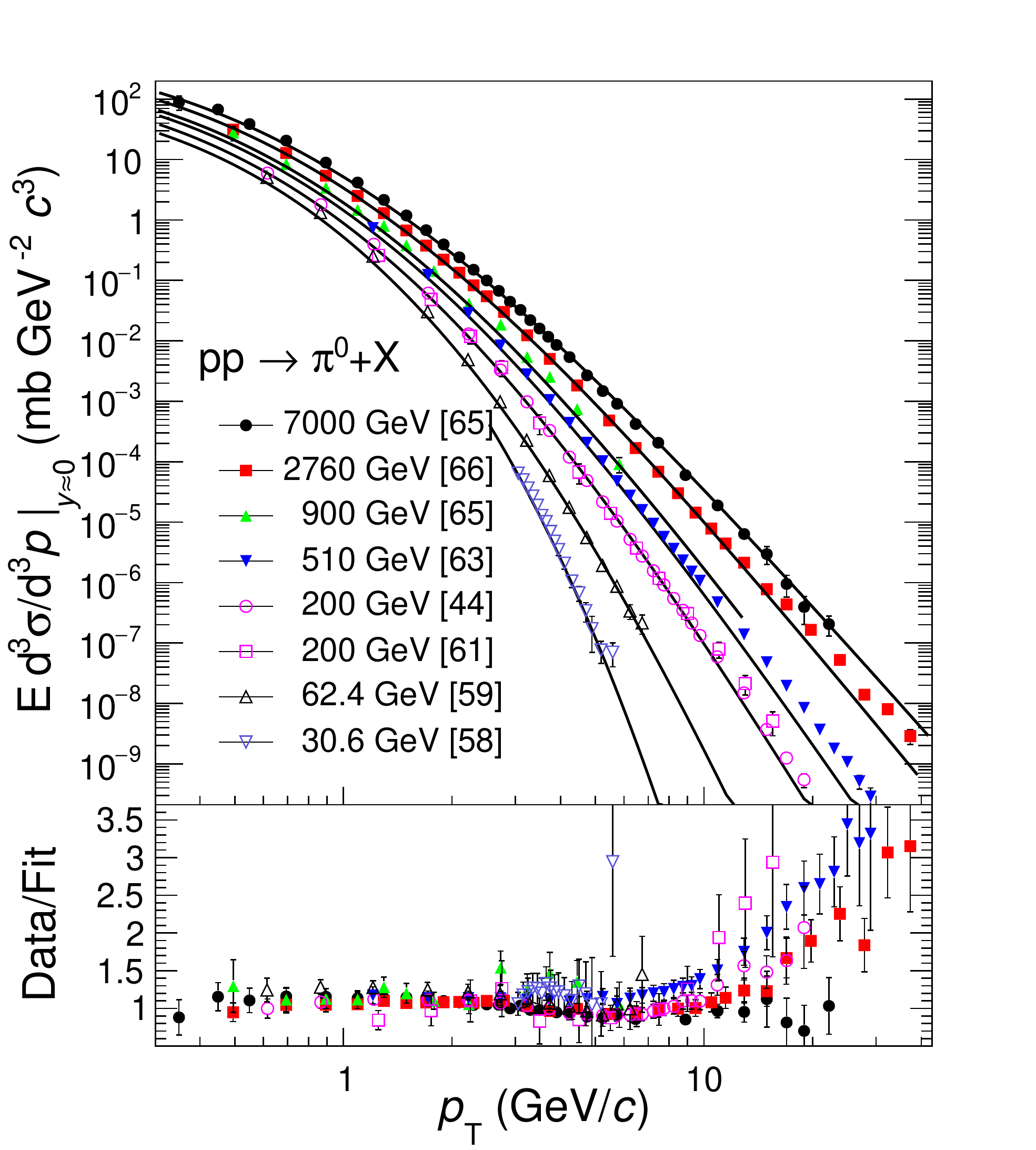}}\!\!\!\!\!\!\!\!\!\!\!
\resizebox{0.53\columnwidth}{!}
{\includegraphics{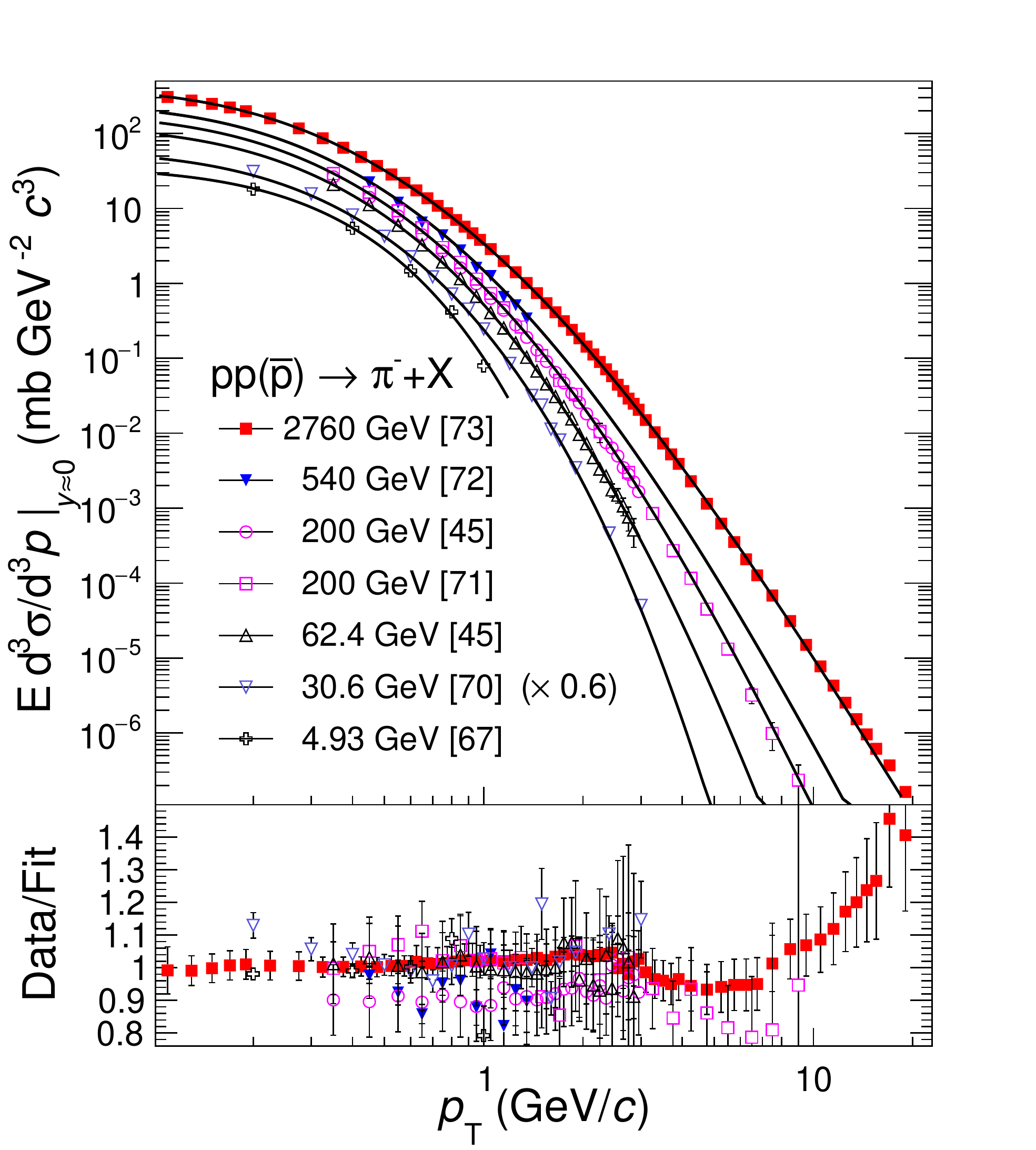}}\!\!\!\!\!\!\!\!\!\!\!
\end{center}
\vskip -2mm
\caption{(color online) Fitting of invariant cross section vs $\pt$ at mid-rapidity and different $\en$ values for 
$\pi^{0}$ (left) and $\pi^{-}$ (right) production in $\pp$ ($\ppbar$ at $\en = 540$~GeV) collisions. Symbols represent the 
data points and the line is the fit function. The $\pi^{-}$ data and line at $\en = 30.6$~GeV are multiplied by 0.6 for a 
better visibility. The ratios data/fit are shown at the bottom.}
\label{pi01}
\end{figure}

\begin{figure}[ht]
\begin{center}
\resizebox{0.53\columnwidth}{!}
{\includegraphics{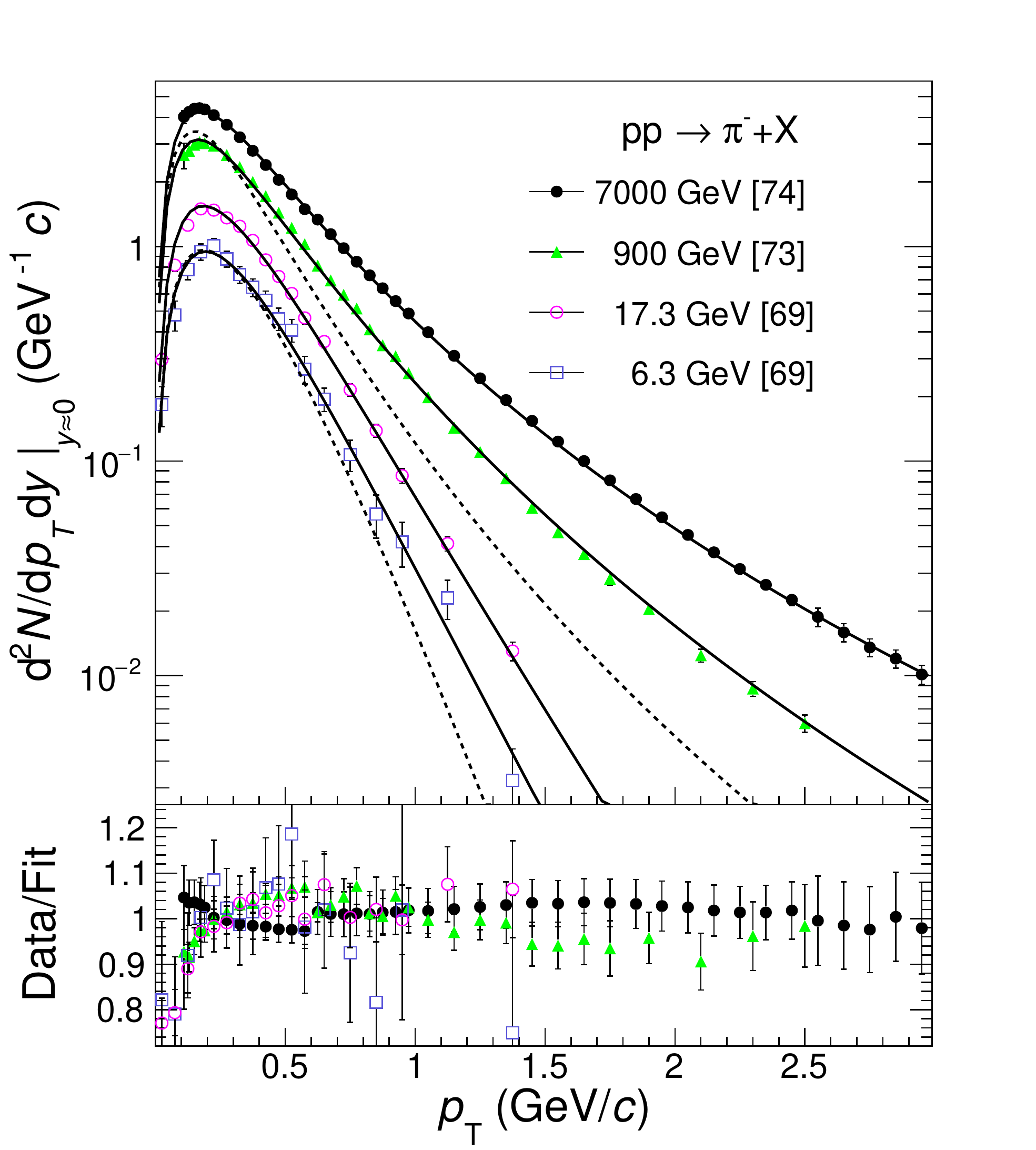}}\!\!\!\!\!\!\!\!\!\!\!
\resizebox{0.53\columnwidth}{!}
{\includegraphics{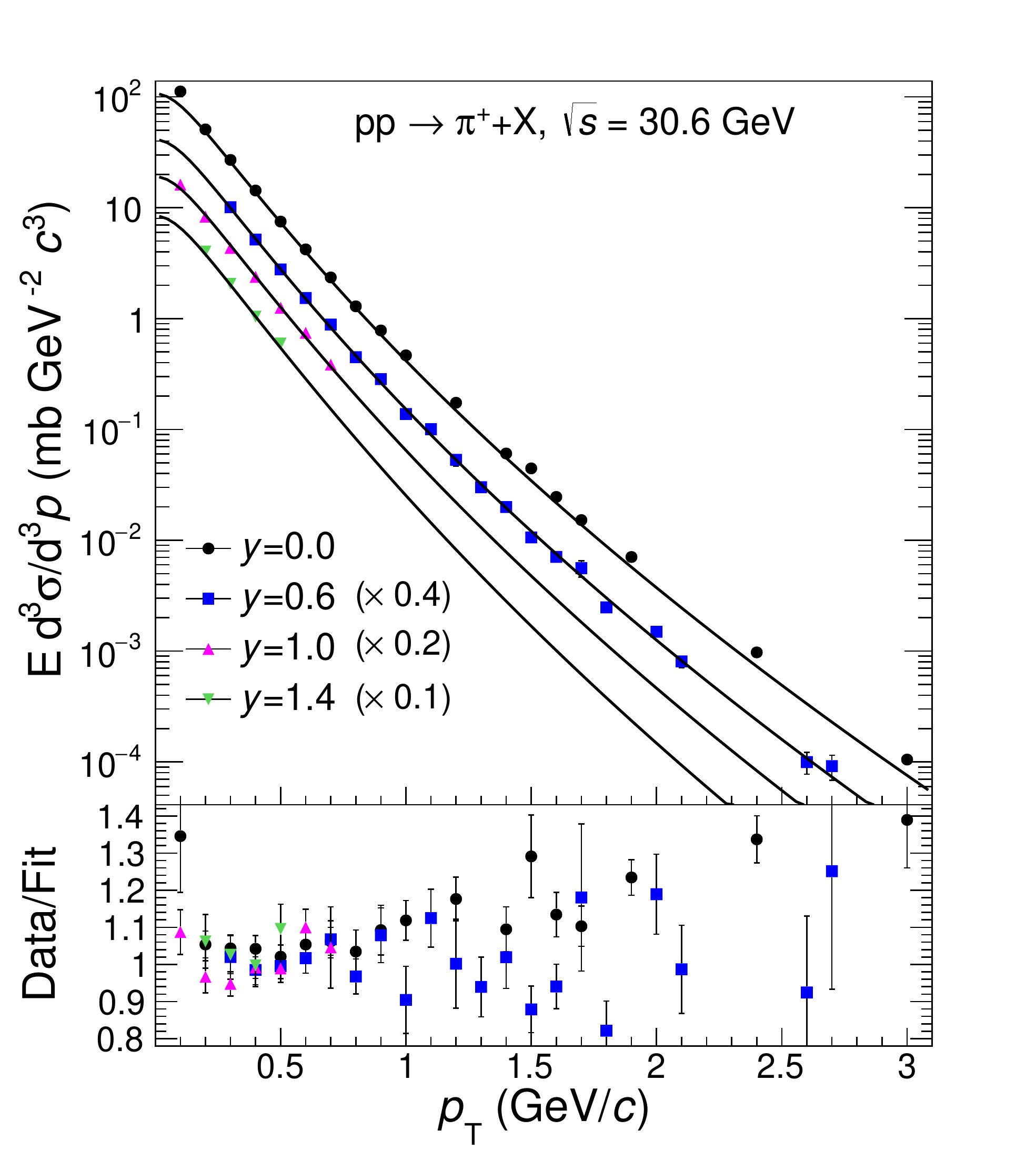}}\!\!\!\!\!\!\!\!\!\!\!
\end{center}
\vskip -2mm
\caption{(color online) Fitting of $\pi^{-}$ invariant yield vs $\pt$ at mid-rapidity and different $\en$ values (left)
and $\pi^{+}$ cross section vs $\pt$ at $\en = 30.6$~GeV and different rapidity values~\cite{Alper} (right).
Data and lines at $y = $ 0.6, 1.0, 1.4 are multiplied by 0.4, 0.2, 0.1, respectively, for a better separation. Dashed lines
show the fit functions at $\en = $ 6.3 and 900~GeV for $v_s = 0$.}
\label{pi231}
\end{figure}

\end{widetext}

\begin{figure}[H]
\begin{center}
\includegraphics[width=1.1\columnwidth]{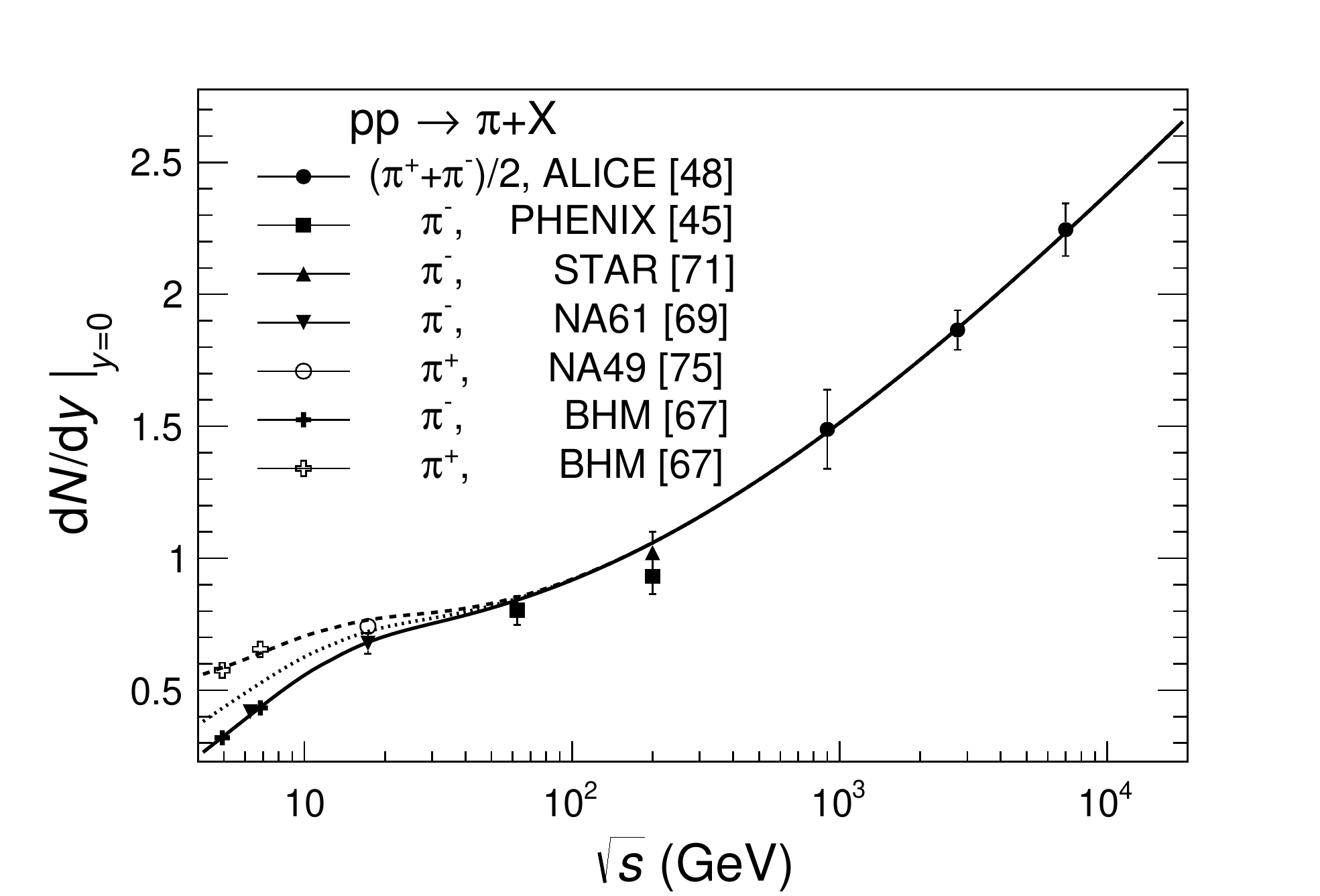}
\end{center}
\vskip -2mm
\caption{Prediction for $\pi^-$ (full line), $\pi^0$ (dotted line) and $\pi^+$ (dashed line) $\pt$-integrated invariant 
yields at mid-rapidity depending on $\en$ and comparison with the available data.}
\label{pidNdy}
\end{figure}

\noindent $\pi^0$~\cite{Phen1,Kourk,Phen3,Phen4,Star2,Star3,Phen5,UA2_1,Alice3,Alice4} and
$\pi^\pm$~\cite{Phen2,Alice2,BHM,Beier76,NA61,Alper,Star4,UA2_2,Alice5,Alice6} inclusive production
$\pt$-spectra measured for different values of $y$ at energies $\en$ from
30.6~GeV~\cite{Kourk} to 7~TeV~\cite{Alice4} for $\pi^0$ and from 4.93~GeV~\cite{BHM} to 7~TeV~\cite{Alice2} for $\pi^\pm$.
The used charged pion data are mostly for $\pi^-$. High-energy data at $\en = $ 0.54, 2.76 and 7~TeV are 
for the averaged ($\pi^- + \pi^+$)/2 production.
From~\cite{Kourk}, we included in the fit only the data obtained with so-called retracted geometry, and from~\cite{Alper},
we included only the data measured at $\en = $ 30.6, 44.6 and 52.8~GeV which cover larger intervals of $\pt$ and $y$.
We did not include in the fit the $\pi^+$ $\pt$-spectra from~\cite{BHM}, which give too large $\chi^2$; however,
our model describes well the corresponding $\pt$-integrated data (see Fig.~\ref{pidNdy}).
Since the charged pion measurement at $\en = 200$~GeV by the STAR Collaboration~\cite{Star4} is for the non-single 
diffractive (NSD) yield, it was converted to an inclusive cross section using $\sigma_{NSD} = 30$~mb~\cite{Star4}.
Figures~\ref{pi01} and~\ref{pi231} (left) show examples of the fits of pion $\pt$-spectra for mid-rapidity and 
different $\en$ values while Fig.~\ref{pi231} (right) shows fits for different values of rapidity at $\en = 30.6$~GeV~\cite{Alper}. 
To demonstrate the quality of the fits, the data points have been divided by the 
corresponding values of the fit function, and the ratios are also plotted.
Generally, the quality is always good. 
Only the $\pi^0$ data~\cite{Star2,Phen5} show a large excess at $\pt > 10$~GeV/$c$.
In Fig.~\ref{pi231} (left), the dashed lines represent the fit functions at $\en = $ 6.3 and 900~GeV for $v_s = 0$
to illustrate the importance of the radial flow in our model.

Note that we did not include in the combined fit the $\pi^{\pm}$ high-statistics data~\cite{NA49} measured in $\pp$ collisions
at $\en = 17.3$~GeV since the published results do not quote
the dominant systematic uncertainties. But we included the $\pi^-$ measurement at the same energy~\cite{NA61}, and
both data sets agree well, as shown in~\cite{NA61}.
We also did not use for the combined fit the charged pion data measured at $\en = 200$~GeV and high rapidities $y = $ 2.95
and 3.3~\cite{Brahms1} but have checked that our model describes them well at $\pt < 2.5$~GeV/$c$. At higher $\pt$,
corresponding to pion energies larger than 25~GeV, the model overestimates the data. So, our model is not valid for such high
rapidities at $\en = 200$~GeV, related to diffractive processes, which is generally expected for a thermal model. Note also that 
the $\pi^\pm$ data~\cite{Alper} provide a large contribution into the $\chi^2$ and $NDF$ values shown in Table~\ref{tab:tab1}. 
A combined fit without these data gives $\chi^2 / NDF = 5959/1734$.

Fig.~\ref{pidNdy} presents an example of our predictions, based on Eqs.~(\ref{eq:d2N})$-$(\ref{eq:mupi}), for the pion 
$\pt$-integrated yields in $\pp$ collisions at mid-rapidity 
and varying $\en$. It shows a good agreement with the available data.

\section{\label{sec5}Quarkonia ($\phi$, $\jpsi$, $\psiprime$, $\upsi$--$\upsipp$)}
\vskip 1mm

\subsection{\label{sec51}$\phi$ meson} 

\begin{widetext}

\begin{figure}[ht]
\begin{center}
\resizebox{0.53\columnwidth}{!}
{\includegraphics{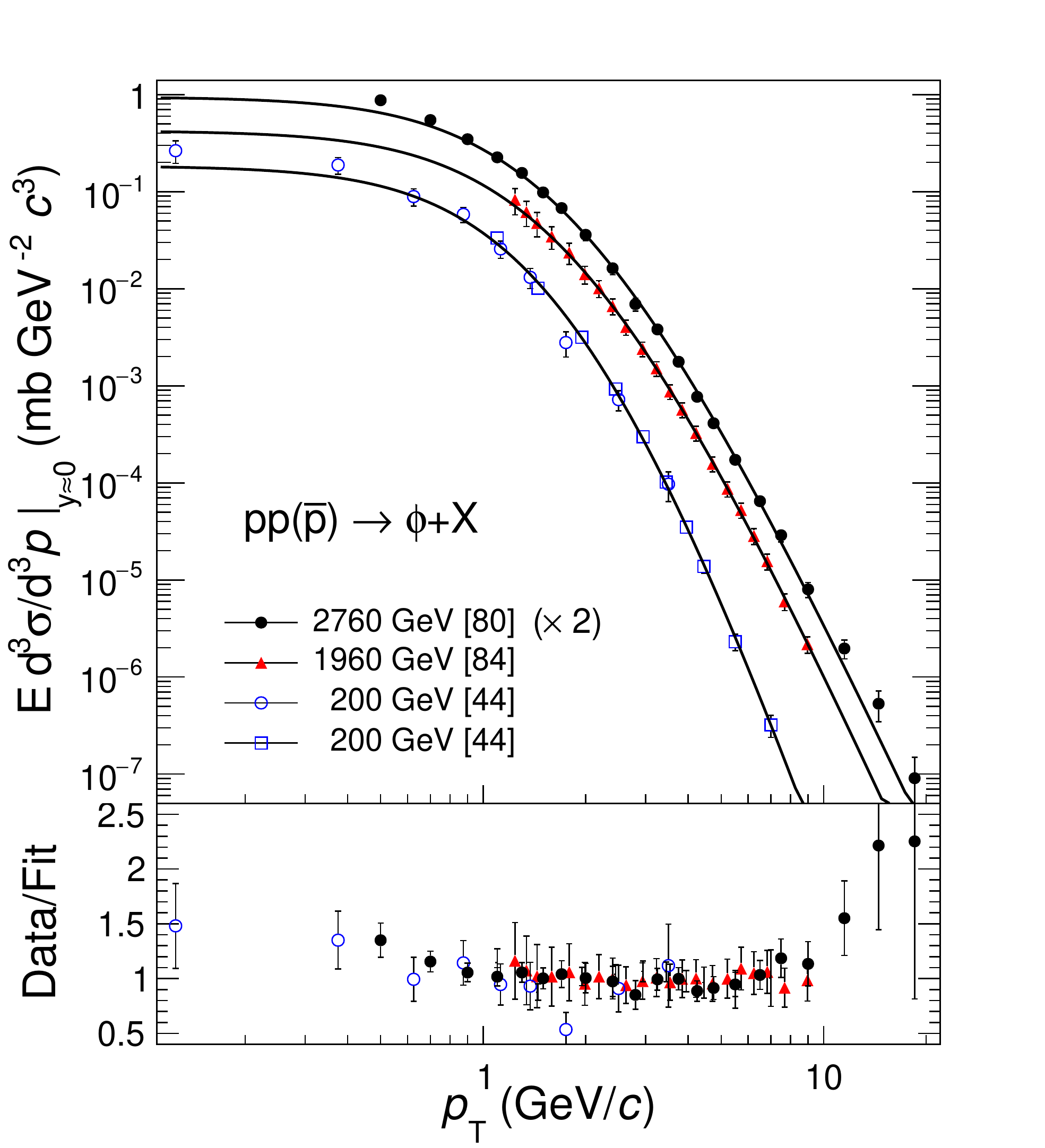}}\!\!\!\!\!\!\!\!\!\!\!
\resizebox{0.53\columnwidth}{!}
{\includegraphics{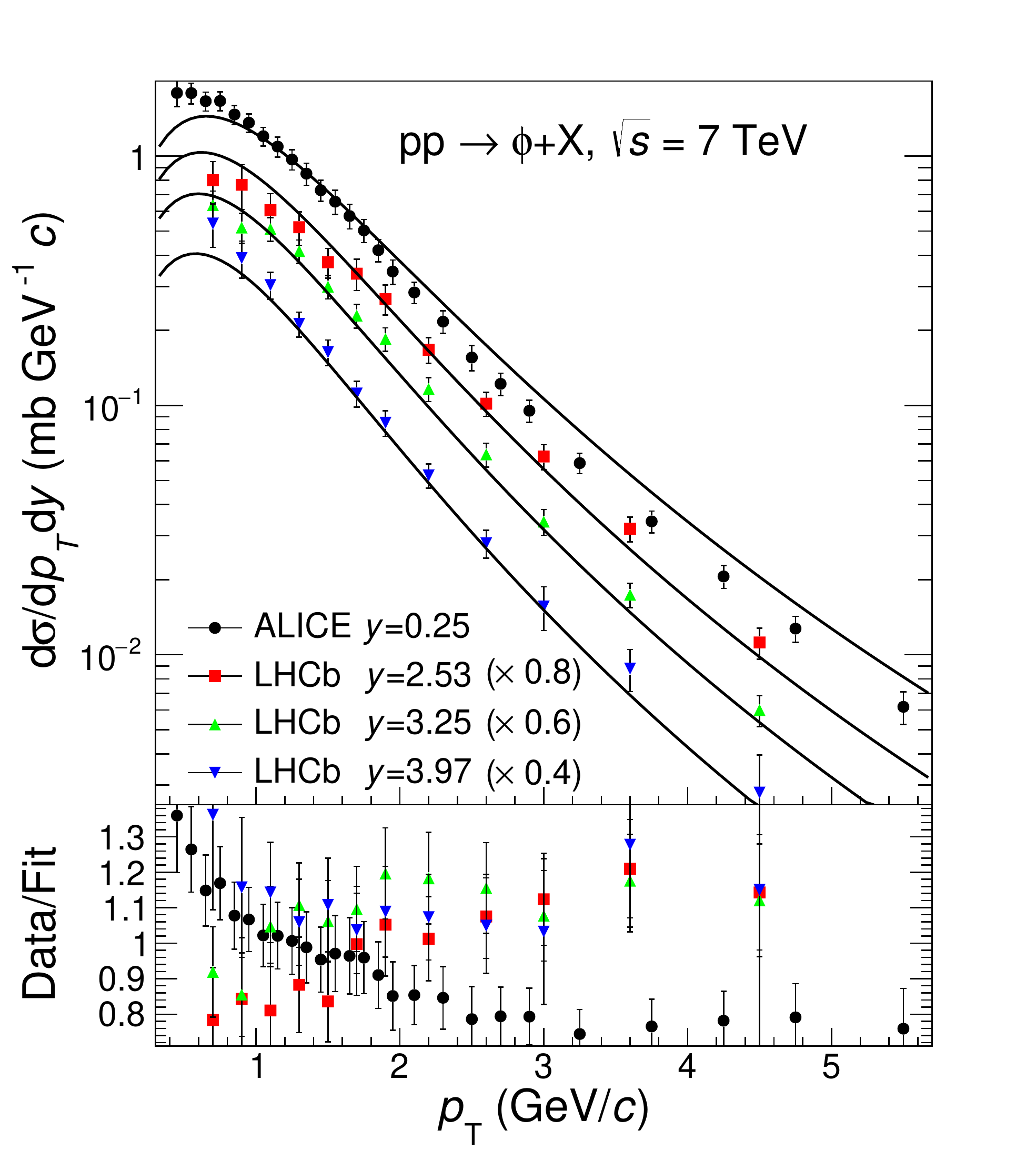}}\!\!\!\!\!\!\!\!\!\!\!
\end{center}
\vskip -5mm
\caption{(color online) Fitting of $\phi$ meson cross section vs $\pt$ at mid-rapidity and different $\en$ values (left)
and at $\en = 7$~TeV and different rapidity values~\cite{LHCb7phi,Alice7phiee} (right).
The data and line for $\en = 2760$~GeV are multiplied by 2 and the ones for $y = $ 2.53, 3.25, 3.97 are multiplied by 0.8, 0.6, 0.4, 
respectively, for a better separation.}
\label{phi12}
\end{figure}
\end{widetext}

\begin{figure}[ht]
\begin{center}
\includegraphics[width=1.1\columnwidth]{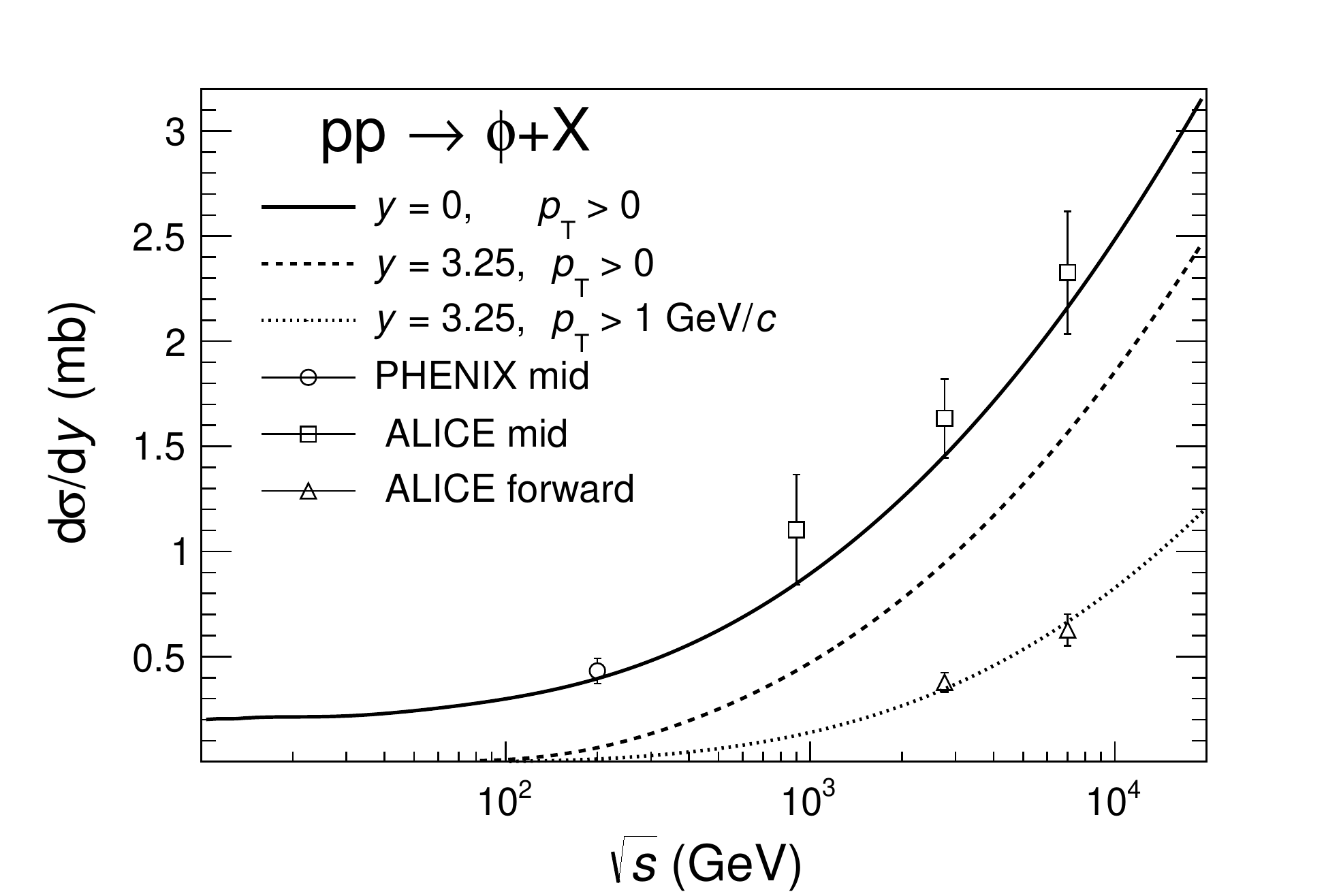}
\end{center}
\vskip -2mm
\caption{$\phi$ meson $\pt$-integrated cross section at mid-rapidity (full line) and forward
rapidity of $y = 3.25$ for $\pt>0$ (dashed line) and $\pt>1$~GeV/$c$ (dotted line) vs $\en$.
Data are from PHENIX~\cite{Phen1}, ALICE~\cite{Alice09phi,Alice3phi,Alice7phiee} for mid-rapidity and
ALICE~\cite{Alice3phi,Alice7phimm} for forward rapidity at $\pt > 1$~GeV/$c$.}
\label{phidsigdy}
\end{figure}

The following results are for the combined fit of $\phi$ meson inclusive production data measured in $\pp$ collisions at 
$\en = $ 17.3~GeV~\cite{NA49phi}, 200~GeV~\cite{Phen1,Phen6}, 900~GeV~\cite{Alice09phi}, 2.76~TeV~\cite{Alice3phi}, 
7~TeV~\cite{LHCb7phi,Alice7phiee,Alice7phimm} and in $\ppbar$ collisions at $\en = $1.96~TeV~\cite{CDFphi}.
The $\pt$-spectrum from~\cite{E735phi} is not included in the fit since its normalization is about six times lower
the one in~\cite{CDFphi} at similar energy.
It appeared that the fitted values of the $\phi$ meson $n$ parameter at different $\en$ are close to the pion ones. 
So, in the parametrization Eq.~(\ref{eq:npar}) for $\phi$, we have fixed some of the parameter values to the ones for pion 
(see Table~\ref{tab:tab1}). Examples of the $\pt$-spectra fits are shown in Fig.~\ref{phi12} for mid-rapidity 
and different $\en$ values (left) and for different values of $y$ at
$\en = 7$~TeV (right).
As an example of our predictions, Fig.~\ref{phidsigdy} presents the $\phi$ meson $\pt$-integrated cross section
in $\pp$ (also $\ppbar$) collisions versus $\en$ at mid-rapidity and

\noindent
at forward rapidity of $y = 3.25$ (for
dimuon decay channel measurements of LHCb~\cite{LHCb7phi} and ALICE~\cite{Alice3phi,Alice7phimm}).
At $y = 3.25$ two values for the $\pt$-integration lower limit are considered: 0 and 1~GeV/$c$.
Comparison of calculations with the available data shows a reasonable agreement.

\subsection{\label{sec52}$\jpsi$ meson} 

The  inclusive $\jpsi$ production consists of prompt component (includes direct production and feed-down from the 
radiative decays of higher charmonium states) and  non-prompt component (includes feed-down from the weak decays of 
bottom hadrons). Fraction of the non-prompt component, denoted usually by $f_B$, is negligible at $\en < $\,100~GeV 
but rises monotonically with $\en$ and $\pt$.
For LHC energies it reaches values about 0.1 at low $\pt$ and larger values at high $\pt$ (see Fig.~\ref{jpsi3}). 
The kinematic distributions of prompt and direct $\jpsi$ are similar and can be described by the same values of 
parameters in Eq.~(\ref{eq:d2N}). Only the normalization constants $\widetilde{V}$ will differ. 
The non-prompt $\jpsi$ has a significantly harder $\pt$ spectrum and narrower $y$ spectrum. Its proper description would require
the use of Eq.~(\ref{eq:d2N}) for the production of bottom mesons and baryons which have several decay channels into $\jpsi$.
To avoid such a complex computation for a rather small fraction of data, we have chosen a simpler approach. 
Namely, for non-prompt $\jpsi$, we use Eq.~(\ref{eq:d2N}) with the same $T$, $n$, $\mu$ and $v_s$ parameters as for 
prompt $\jpsi$. To describe the harder $\pt$-spectrum of non-prompt $\jpsi$, 
we assume that the mass in $\mt$ in the
\begin{widetext}

\begin{figure}[H]
\begin{center}
\resizebox{0.53\columnwidth}{!}
{\includegraphics{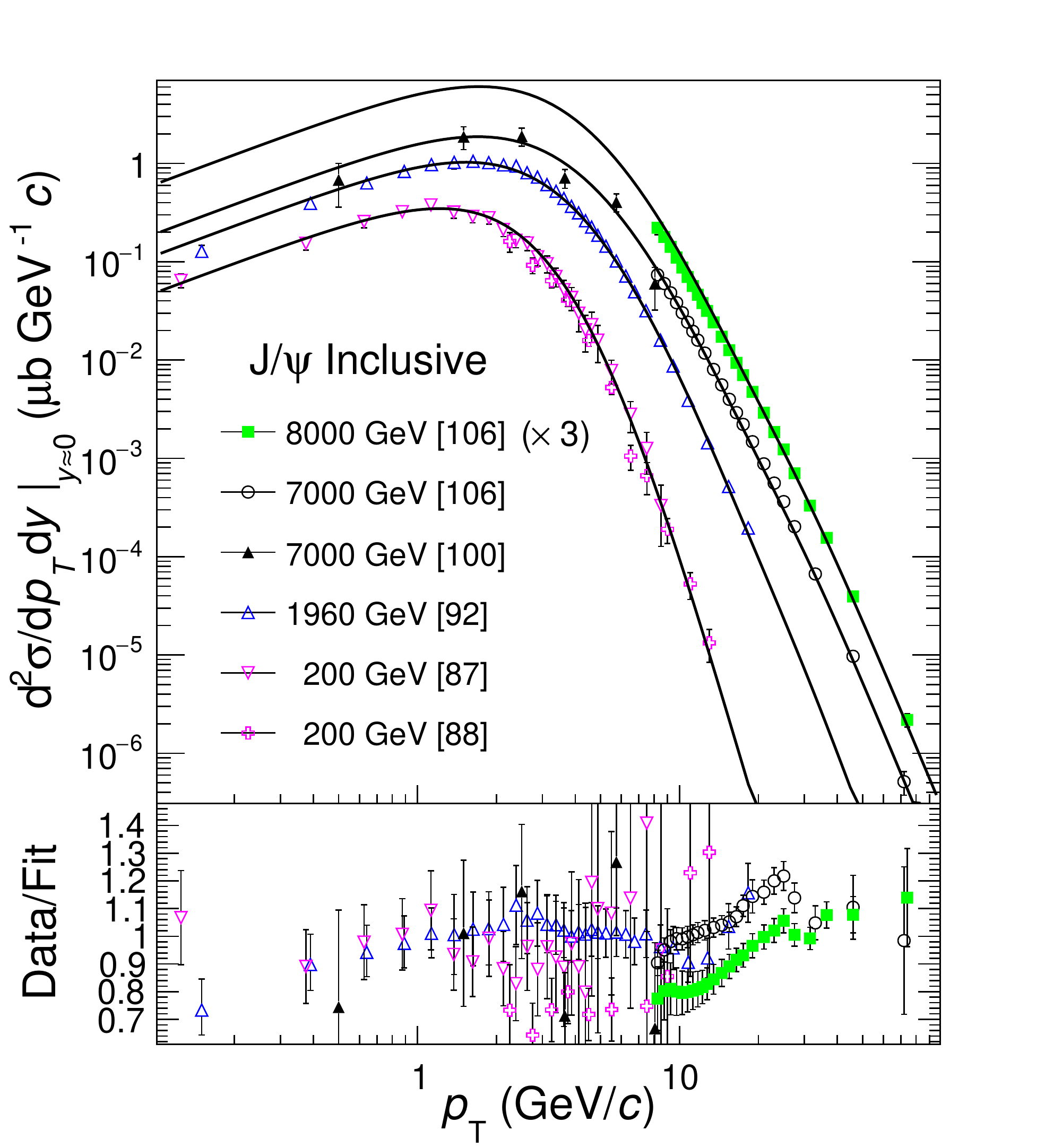}}\!\!\!\!\!\!\!\!\!\!\!
\resizebox{0.53\columnwidth}{!}
{\includegraphics{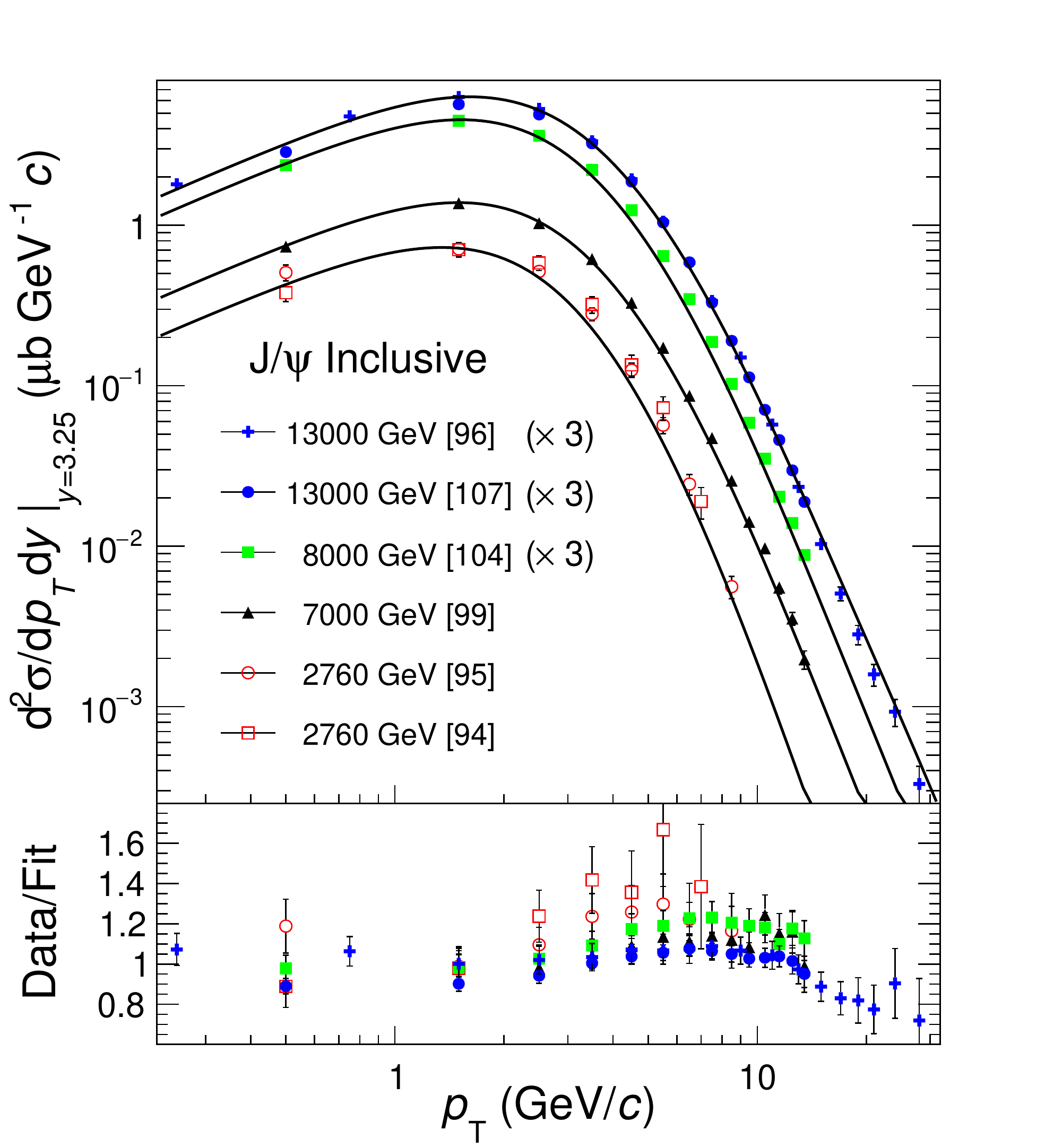}}\!\!\!\!\!\!\!\!\!\!\!
\end{center}
\vskip -2mm
\caption{(color online) Fitting of inclusive $\jpsi$ meson cross section vs $\pt$ at different $\en$ values
for mid-rapidity (left) and forward rapidity of $y = 3.25$ (right). Data and lines at $\en = $13, 8~TeV 
are multiplied by 3 for a better visibility.}
\label{jpsi01}
\end{figure}
\end{widetext}

\noindent 
corresponding Eq.~(\ref{eq:d2N}) is larger the $\jpsi$ mass by some factor $c_m$ and the normalization grows with $\pt$ 
according to the parametrization
\be
\widetilde{V} = \widetilde{V}_{NP} (1 + \frac{c_1}{1 + (c_2 / \pt)^4}) \, ,
\label{eq:NPjpsi} 
\ee
where $\widetilde{V}_{NP}$, $c_1$, $c_2$ and $c_m$ are fit parameters. 
Also, to ensure the narrowness of the non-prompt $y$-spectrum, 
we multiply the corresponding $\eta_{max}$ in Eq.~(\ref{eq:etav}) by another fit parameter $c_{\eta} < 1$.
We have performed a combined fit of the available prompt and non-prompt or inclusive $\jpsi$ production 
$\pt$-spectra~\cite{NA3,Phen7,Star5,CDFpsi1,D0jpsi,CDFjpsi1,CDFjpsi2,CMS3QQ,Alice3jpsi,LHCb3jpsi,Alice513psi,CMS7jpsi,
Atlas7jpsi,LHCb7jpsi,Alice7jpsi,CMS7psi1,Alice7QQ,CMS7psi2,LHCb8QQ,Alice8QQ,Atlas78psi,LHCb13jpsi}
measured at energies $\en$ from 19.4~GeV~\cite{NA3} to 13~TeV~\cite{Alice513psi,LHCb13jpsi} in $\pp$ collisions and at
1.8~TeV~\cite{CDFpsi1,D0jpsi,CDFjpsi1} and 1.96~TeV~\cite{CDFjpsi2} in $\ppbar$ collisions. 
The $\pt$-spectrum from~\cite{UA1jpsi} is not included in the fit since its normalization is a factor of 2.5 lower
than expected within our model, which however describes well the shape of this spectrum.
The fit gives, in addition to the values for the model parameters, $\chi^2$ and $NDF$ listed in 
Table~\ref{tab:tab1}, the following values for the non-prompt component parameters:
$\widetilde{V}_{NP} = 82.1\pm0.7~{\rm GeV}^{-3},\, c_1=2.1,\, c_2=26.3~{\rm GeV},\, c_m = 1.4,\, c_{\eta} = 0.82$.

We illustrate then some results of the fit.
Inclusive $\jpsi$ $\pt$-spectra at different $\en$ values are shown in Fig.~\ref{jpsi01} 
for midrapidity (left) and forward rapidity of $y = 3.25$
\begin{figure}[ht]
\begin{center}
\includegraphics[width=1.1\columnwidth]{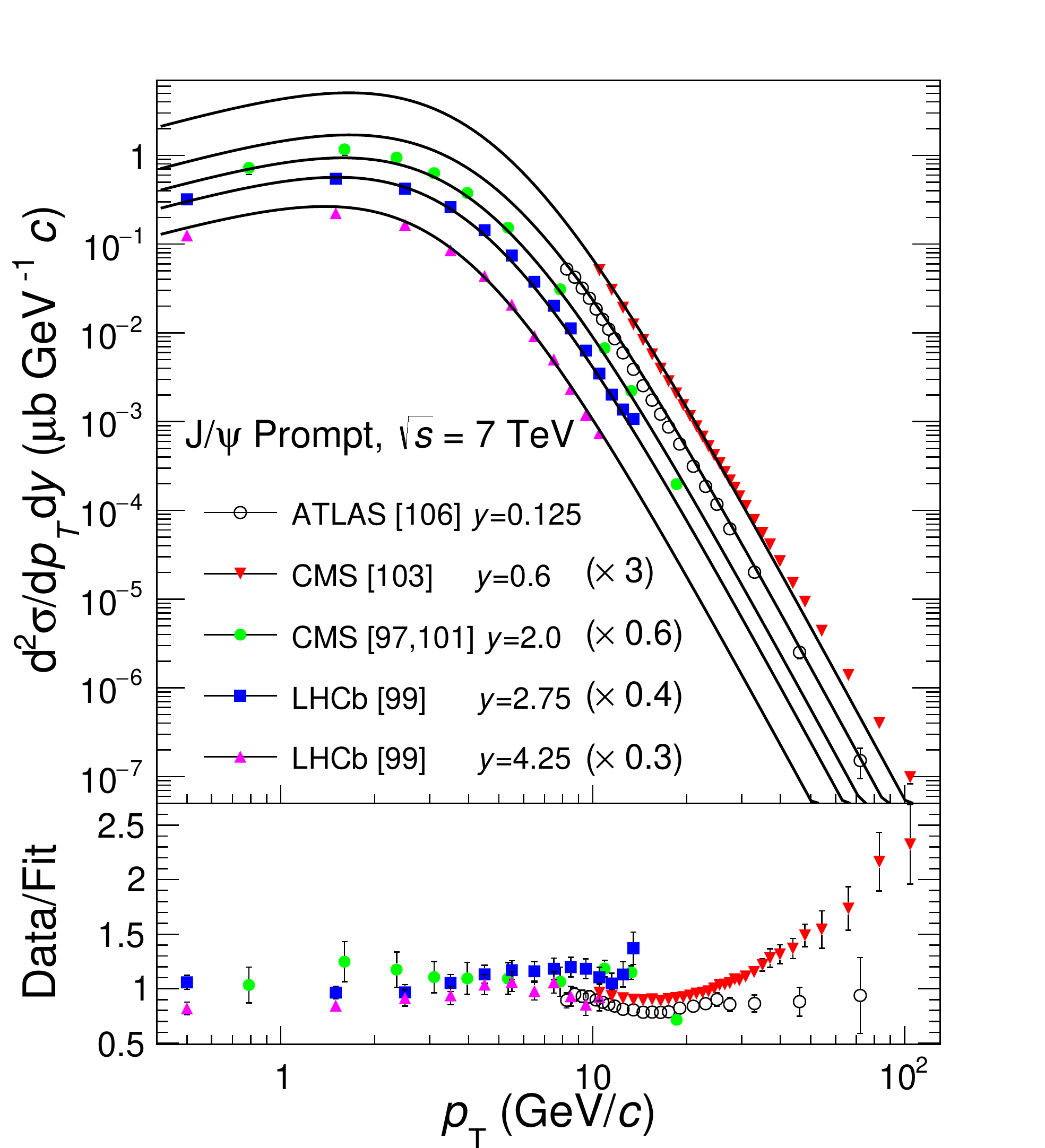}
\end{center}
\vskip -2mm
\caption{(color online) Fitting of prompt $\jpsi$ meson cross section vs $\pt$ at $\en = 7$~TeV and 
different rapidity values. Data and lines at $y = $ 0.6, 2.0, 2.75, 4.25 are multiplied by 3, 0.6, 0.4, 0.3, 
respectively, for a better separation.}
\label{jpsi2}
\end{figure}

\noindent 
(right). Data points of ATLAS~\cite{Atlas78psi} in Fig.~\ref{jpsi01} correspond to $0.25<|y|<0.5$.
Figure~\ref{jpsi2} shows the prompt $\jpsi$ $\pt$ spectra for different rapidity values at $\en = 7$~TeV. 
In Fig.~\ref{jpsi3}, our predictions for the $\pt$ dependence of the $\jpsi$ non-prompt fraction at different $\en$ 
are compared with existing published data.
We demonstrate also a good agreement of our predictions with the available data on the inclusive $\jpsi$ meson 
$\pt$-integrated cross section (for $\pt>0$) as a function of $y$ at three $\en$ values (Fig.~\ref{jpsiy}) 
and as a function of $\en$ at mid-rapidity and forward rapidity (Fig.~\ref{jpsiy0}).
\begin{figure}[ht]
\begin{center}
\includegraphics[width=1.1\columnwidth]{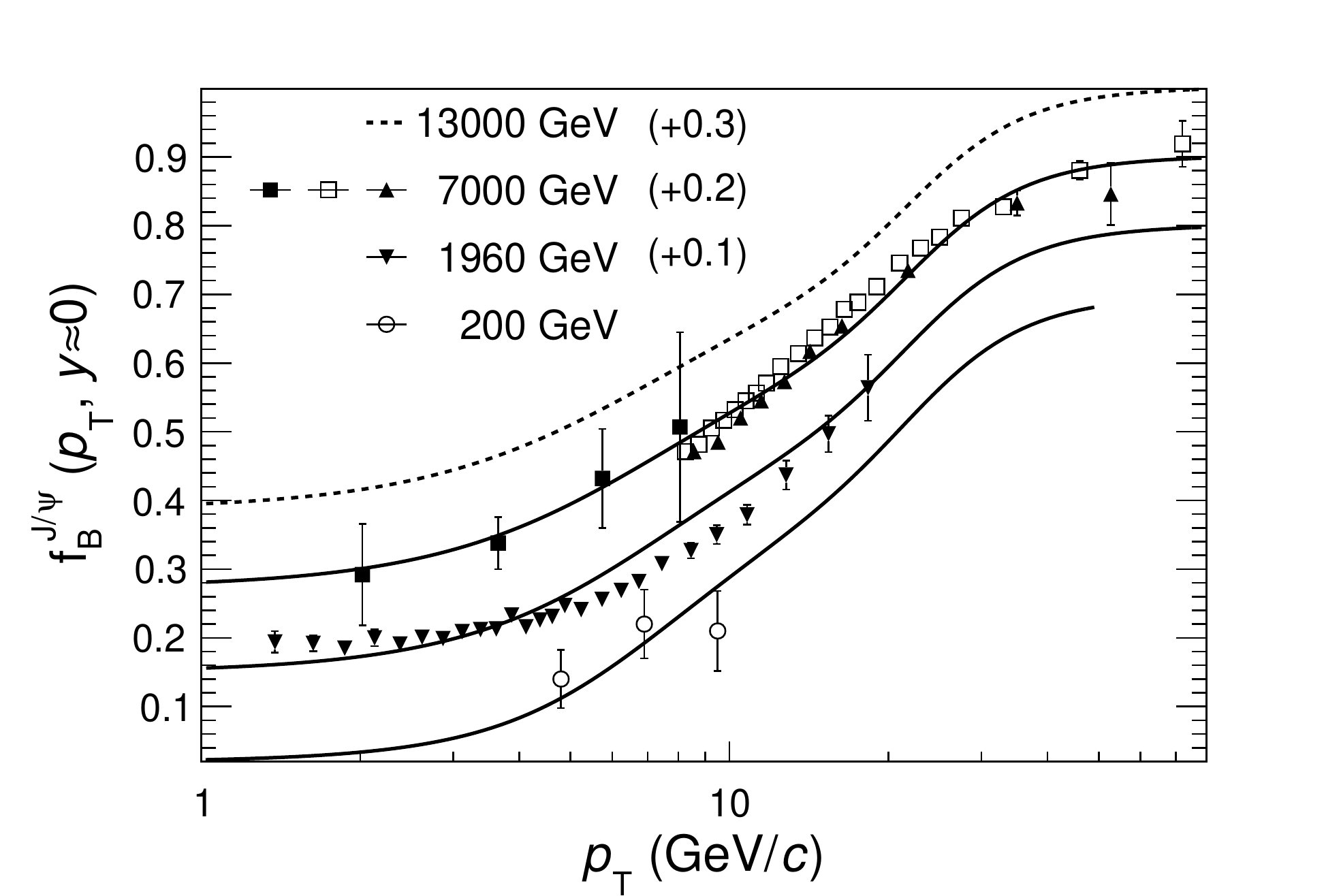}
\end{center}
\vskip -2mm
\caption{Fraction of non-prompt $\jpsi$ production versus $\pt$ at mid-rapidity and different $\en$. Data and 
lines at $\en = $ 13~TeV, 7~TeV~\cite{Alice7jpsi,CMS7psi1,Atlas78psi}, 1.96~TeV~\cite{CDFjpsi2}, 0.2~TeV~\cite{Star5}
are shifted up by 0.3, 0.2, 0.1, 0, respectively, for a better separation.}
\label{jpsi3}
\end{figure}

\begin{figure}[ht]
\begin{center}
\includegraphics[width=1.1\columnwidth]{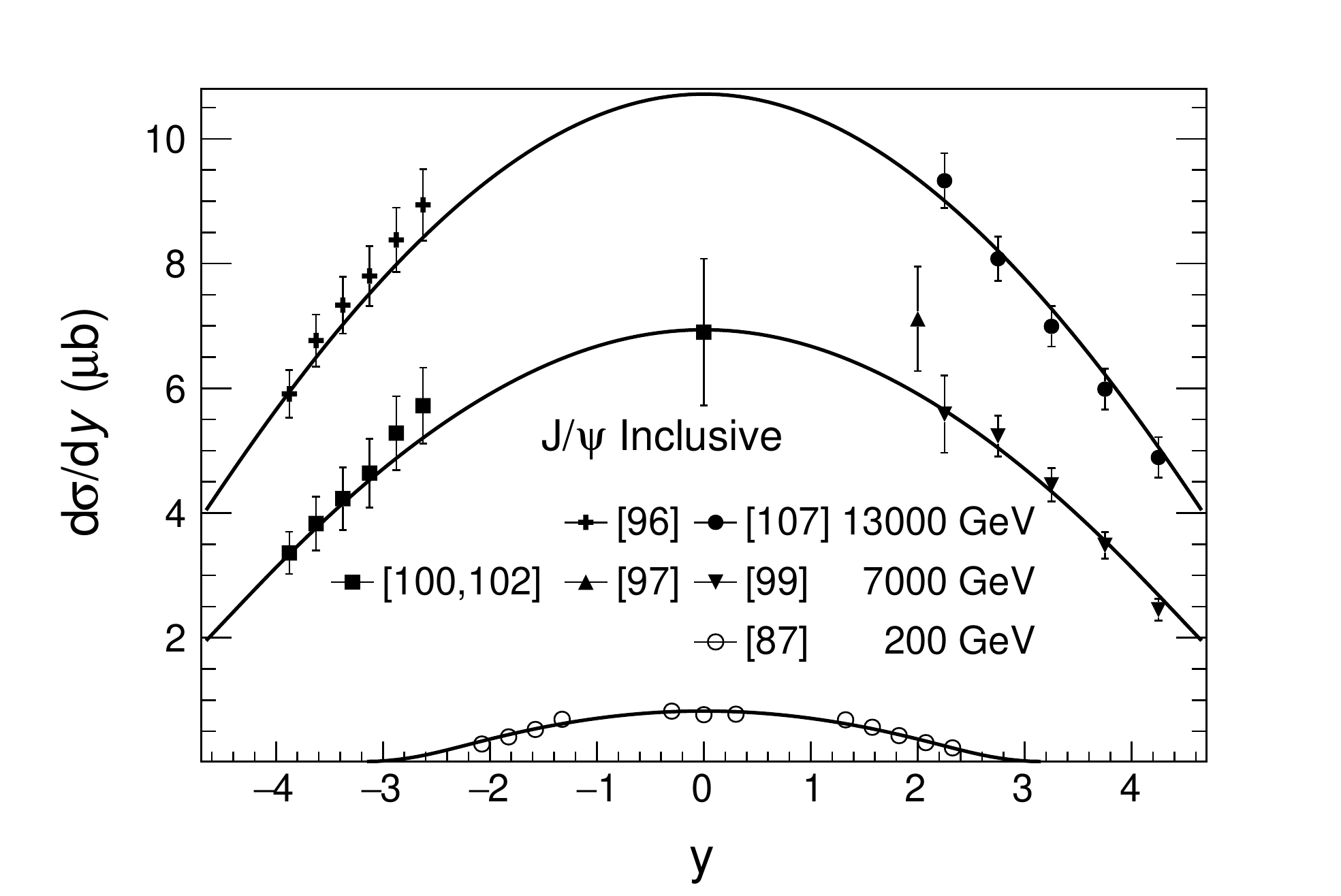}
\end{center}
\vskip -2mm
\caption{Inclusive $\jpsi$ meson $\pt$-integrated cross section as a function of $y$ at different $\en$ 
and comparison with the data.}
\label{jpsiy}
\end{figure}

\begin{figure}[ht]
\begin{center}
\includegraphics[width=1.1\columnwidth]{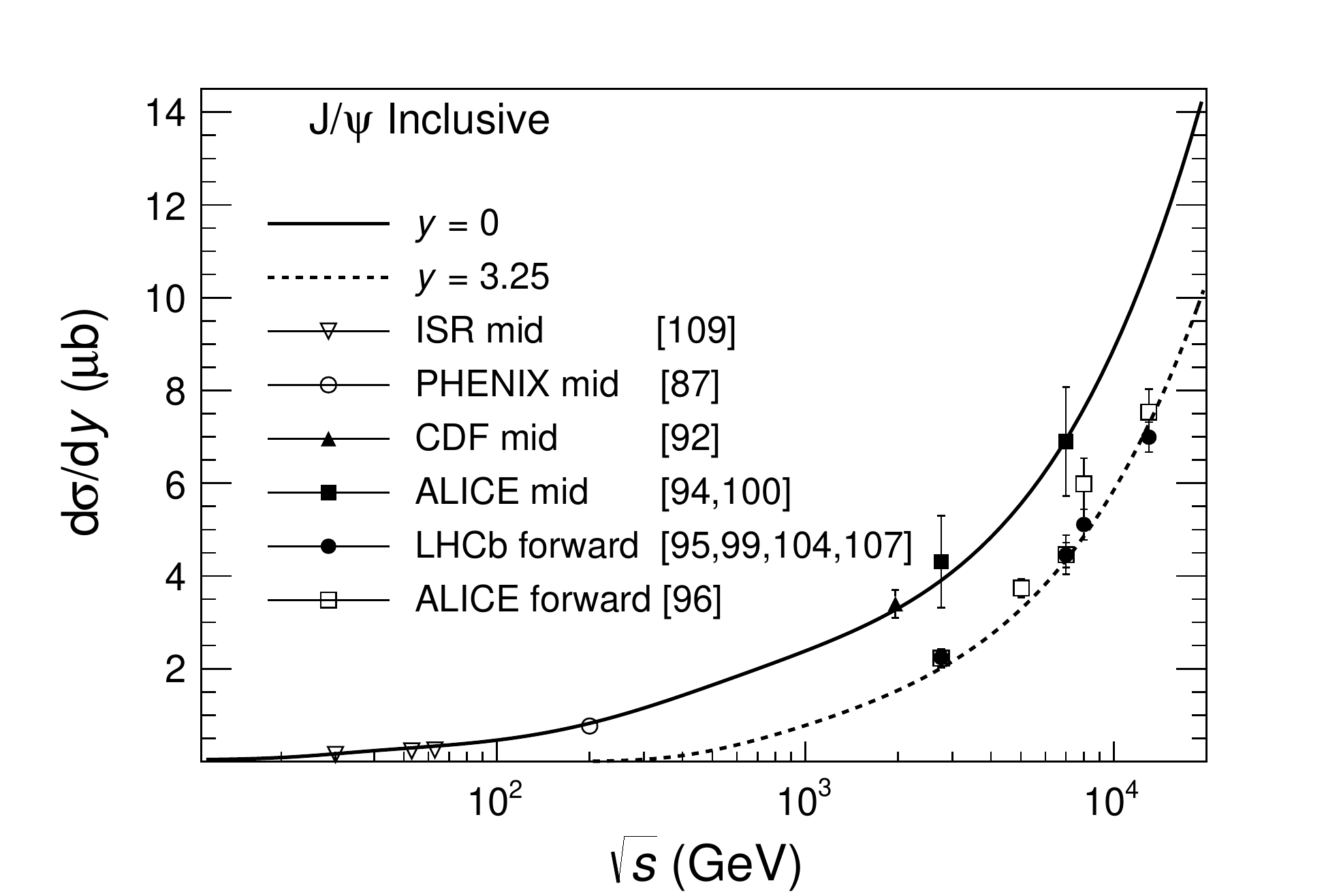}
\end{center}
\vskip -5mm
\caption{Inclusive $\jpsi$ meson $\pt$-integrated cross section as a function of $\en$ at mid-rapidity (full line) 
and forward rapidity of $y = 3.25$ (dashed line) and comparison with the data.}
\label{jpsiy0}
\end{figure}

\subsection{\label{sec53}$\psiprime$ meson} 

\begin{widetext}

\begin{figure}[ht]
\begin{center}
\resizebox{0.53\columnwidth}{!}
{\includegraphics{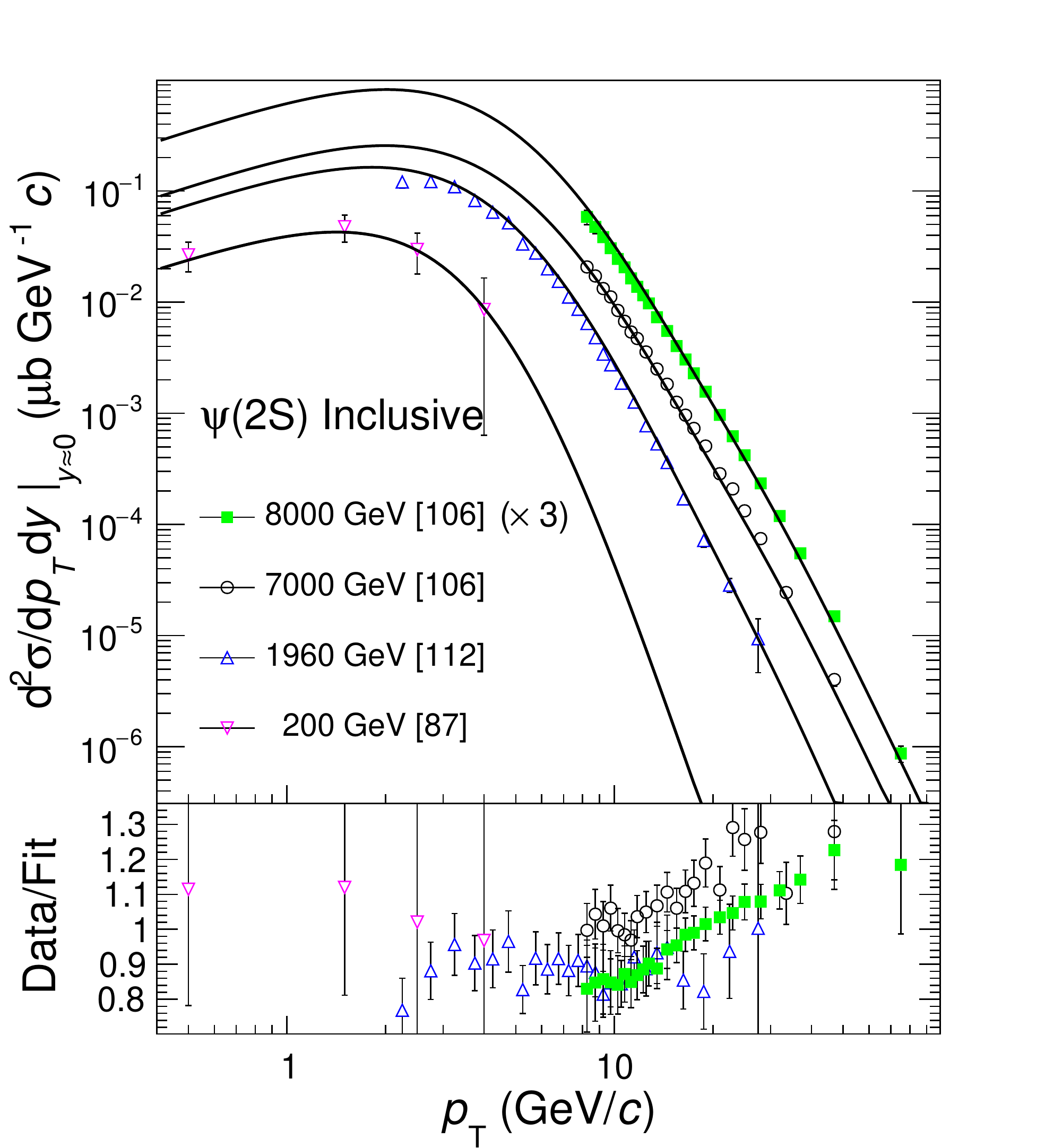}}\!\!\!\!\!\!\!\!\!\!\!
\resizebox{0.53\columnwidth}{!}
{\includegraphics{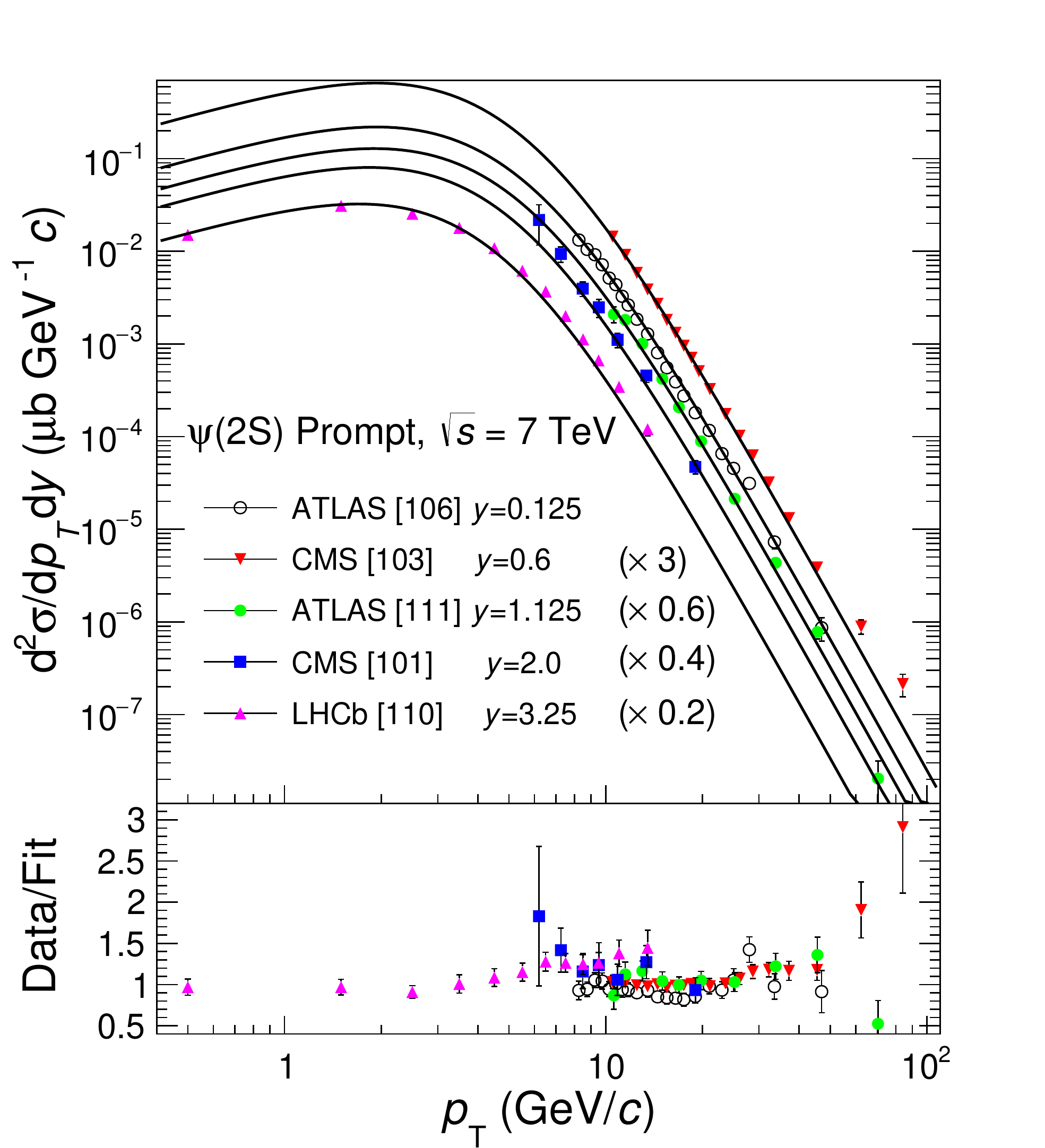}}\!\!\!\!\!\!\!\!\!\!\!
\end{center}
\vskip -4mm
\caption{(color online) Fitting of the $\psiprime$ meson cross section vs $\pt$ at mid-rapidity and 
different $\en$ values (left) and at $\en = 7$~TeV and different rapidity values (right). Data and lines 
are multiplied by the numbers, indicated in the parentheses, for a better separation.}
\label{psi01}
\end{figure}
\end{widetext}

As $\psiprime$ is a charmonium state, similar to $\jpsi$, its production prompt and non-prompt components
can be described similarly using Eqs.~(\ref{eq:d2N}) and~(\ref{eq:NPjpsi}) (where $m$ now is the $\psiprime$ mass)
and parameters $c_m,\, c_{\eta}$. 
We have performed a combined fit of the available prompt and non-prompt or inclusive $\psiprime$ production data
measured in $\pp$ collisions at $\en = $ 200~GeV~\cite{Phen7},
7--13~TeV~\cite{Alice513psi,CMS7psi1,Alice7QQ,CMS7psi2,Alice8QQ,Atlas78psi,LHCb7psi,Atlas7psi} 
and in $\ppbar$ collisions at $\en = $ 1.8~TeV~\cite{CDFpsi1}, 1.96~TeV~\cite{CDFpsi2}.
The resulting fit parameter values are listed in Table~\ref{tab:tab1} (some of them are fixed to the corresponding
values of $\jpsi$). 
Additional parameters for the non-prompt $\psiprime$ are:
$\widetilde{V}_{NP} = 15.9\pm0.1~{\rm GeV}^{-3},\, c_m = 1.3$,\, and $c_1,\, c_2,\, c_{\eta}$ coincide with the ones of $\jpsi$.
Examples of the $\pt$-spectra fits are shown in Fig.~\ref{psi01} for mid-rapidity and different $\en$ 
values (left) and for different values of $y$ at $\en = 7$~TeV (right). 
Data points of ATLAS~\cite{Atlas78psi} in the left panel correspond to

\begin{figure}[H]
\begin{center}
\includegraphics[width=1.1\columnwidth]{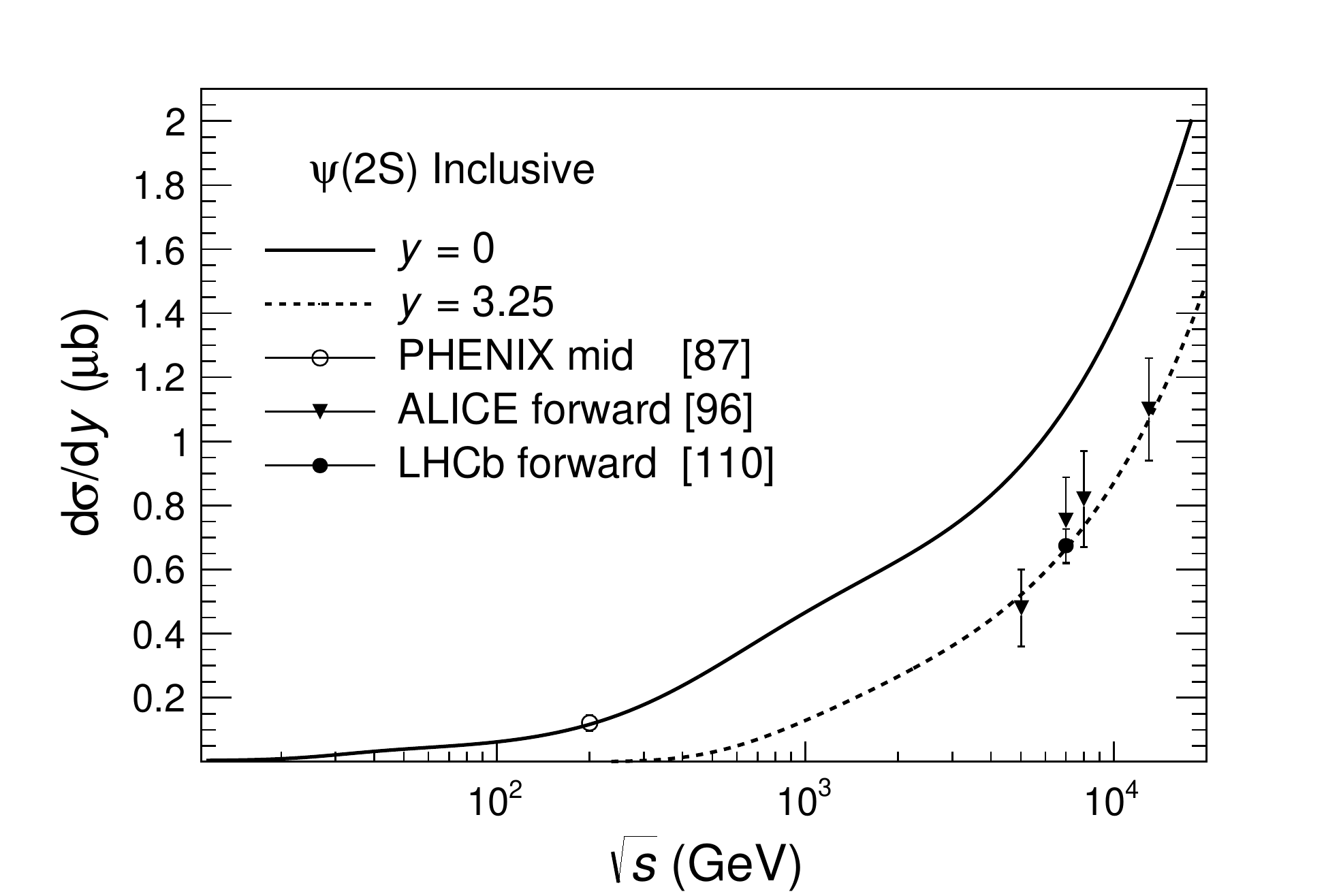}
\end{center}
\vskip -5mm
\caption{Inclusive $\psiprime$ meson $\pt$-integrated cross section as a function of $\en$ at mid-rapidity 
(full line) and forward rapidity of $y = 3.25$ (dashed line) and comparison with the data.}
\label{psiy0}
\end{figure}

\noindent 
$0.25<|y|<0.5$. As an example of our predictions, Fig.~\ref{psiy0} presents the inclusive $\psiprime$ meson $\pt$-integrated
(for $\pt>0$) cross section in $\pp$ (also high energy $\ppbar$) collisions versus $\en$ at mid-rapidity and at forward 
rapidity of $y = 3.25$. Comparison of calculations with the available data shows a reasonable agreement. 

\subsection{\label{sec54}$\upsi$, $\upsip$, $\upsipp$ mesons}

Here, we discuss the fits of $\ups$-family mesons inclusive production data~\cite{CMS3QQ,Alice7QQ,LHCb8QQ,Alice8QQ,
E866upsi,Phen8,Star6,Phen9,CDFupsi,D0upsi,LHCb3upsi,LHCb7upsi,Atlas7upsi,CMS7upsi1,CMS7upsi2,LHCb78upsi} 
measured at energies $\en$ from 38.8~GeV~\cite{E866upsi} to 8~TeV~\cite{LHCb8QQ,Alice8QQ,LHCb78upsi} in $\pp$ collisions and at
1.8~TeV~\cite{CDFupsi} and 1.96~TeV~\cite{D0upsi} in $\ppbar$ collisions.
First, a combined fit of more copious $\upsi$ data was done, and the resulting parameter values are given in Table~\ref{tab:tab1}.
Then, separate combined fits were performed for $\upsip$ and $\upsipp$ data with free parameters $\widetilde{V}$ and $p_3$--$p_6$,
fixing all other parameters to the corresponding values of the $\upsi$ fit. The results for $\upsip$ ($\upsipp$) are
$\widetilde{V} = 0.093\pm0.001\, (0.034\pm0.001)~{\rm GeV}^{-3},\; p_3=0.6\, (0.4),\; p_4=7.5\, (5.8),\; p_5=-8.0\, (-8.0),\; 
p_6=0.0548\, (0.0531),\; \chi^2 / NDF = 3443/491\, (3005/478)$. Note that such large ratios $\chi^2 / NDF$ for $\ups$ family 
are mostly due to the somewhat poor match between results of different LHC experiments. Moreover, two different measurements of
the LHCb Collaboration~\cite{LHCb7upsi,LHCb78upsi} at
\begin{widetext}

\begin{figure}[ht]
\begin{center}
\resizebox{0.53\columnwidth}{!}
{\includegraphics{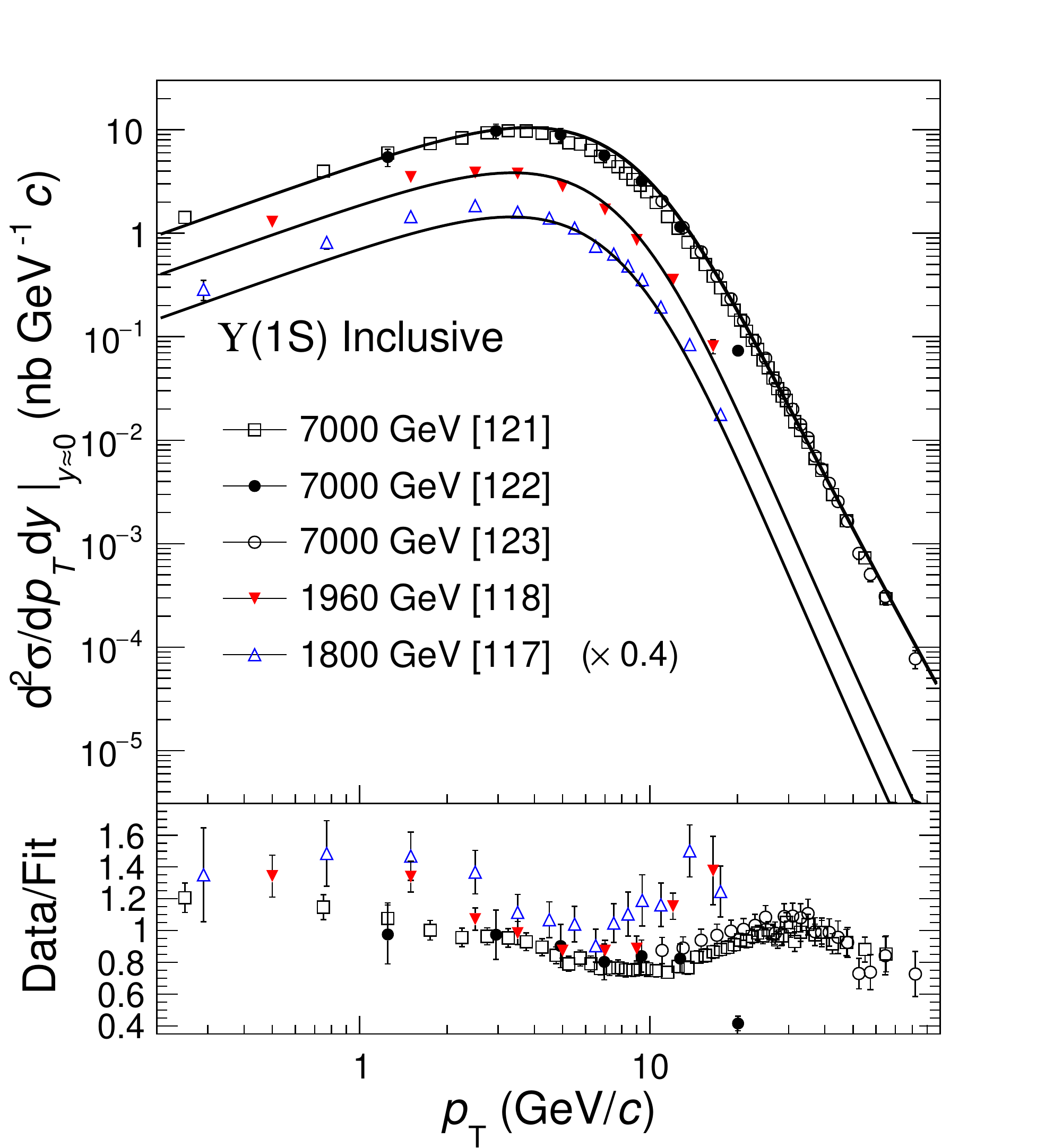}}\!\!\!\!\!\!\!\!\!\!\! 
\resizebox{0.53\columnwidth}{!}
{\includegraphics{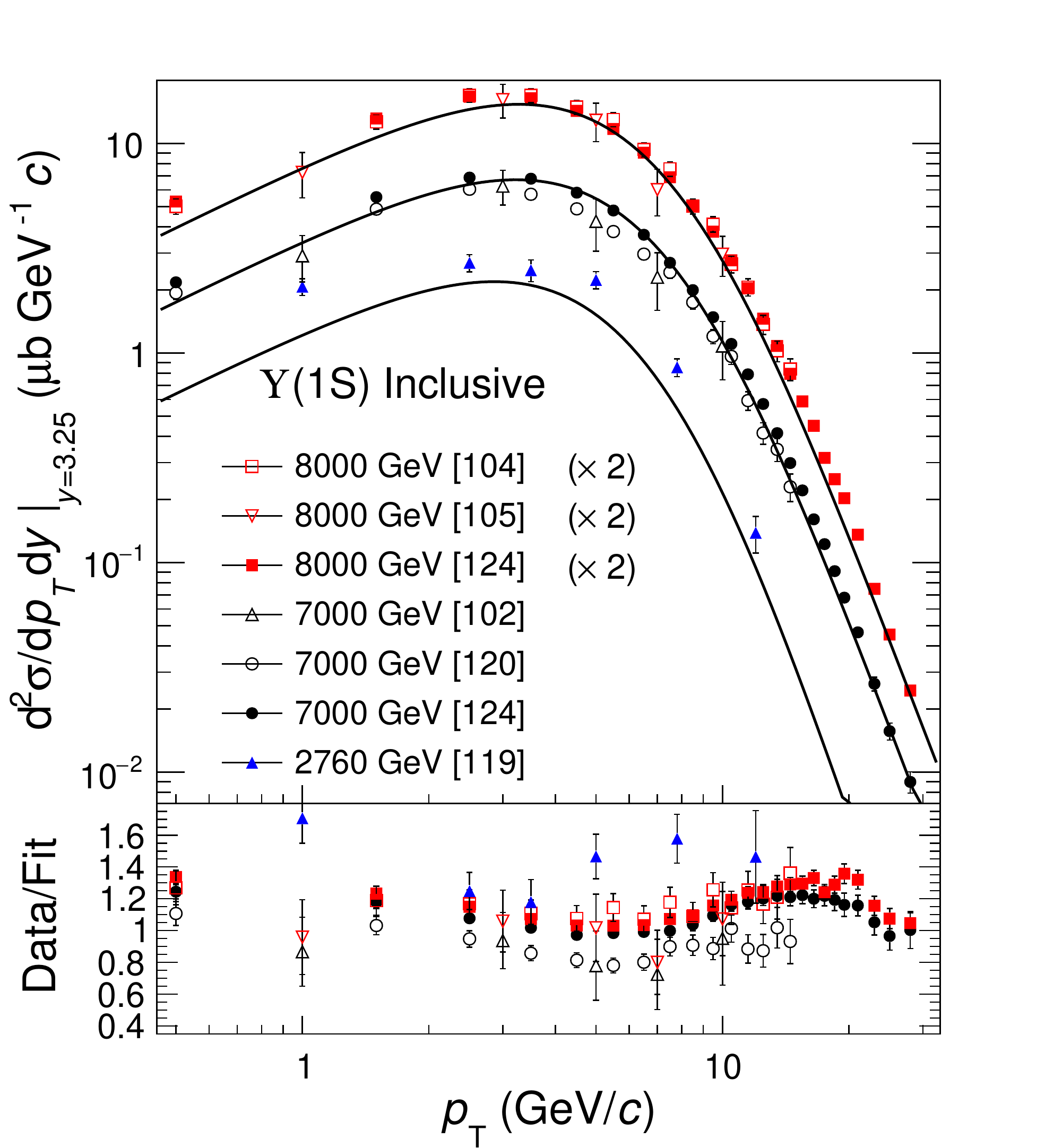}}\!\!\!\!\!\!\!\!\!\!\!
\end{center}
\vskip -2mm
\caption{(color online) Fitting of $\upsi$ meson cross section vs $\pt$ at different $\en$ values
for mid-rapidity (left) and forward rapidity of $y = 3.25$ (right). Data and lines at $\en = $ 1.8, 8~TeV 
are multiplied by 0.4, 2, respectively, for a better visibility.}
\label{upsi01}
\end{figure}
\end{widetext}

\begin{figure}[ht]
\begin{center}
\includegraphics[width=1.1\columnwidth]{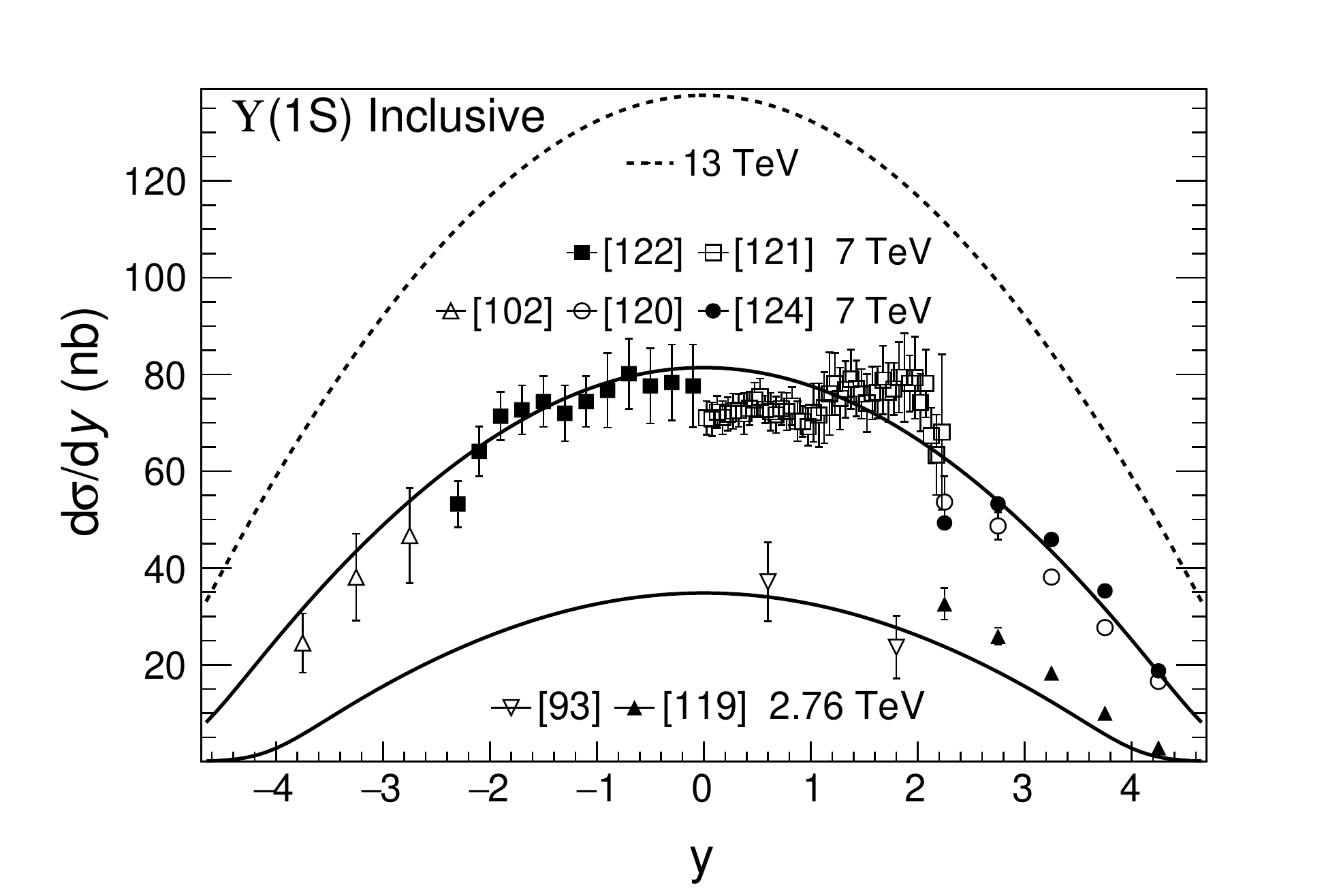}
\end{center}
\vskip -2mm
\caption{$\upsi$ meson $\pt$-integrated cross section as a function of $y$ at different $\en$ values
and comparison with the data.}
\label{upsiy}
\end{figure}

\noindent 
$\en = 7$~TeV do not agree well, and the data at $\en = 2.76$~TeV~\cite{LHCb3upsi} seem too high with respect to 
the model predictions (see Figs.~\ref{upsi01} and~\ref{upsiy}).

\begin{figure}[ht]
\begin{center}
\includegraphics[width=1.1\columnwidth]{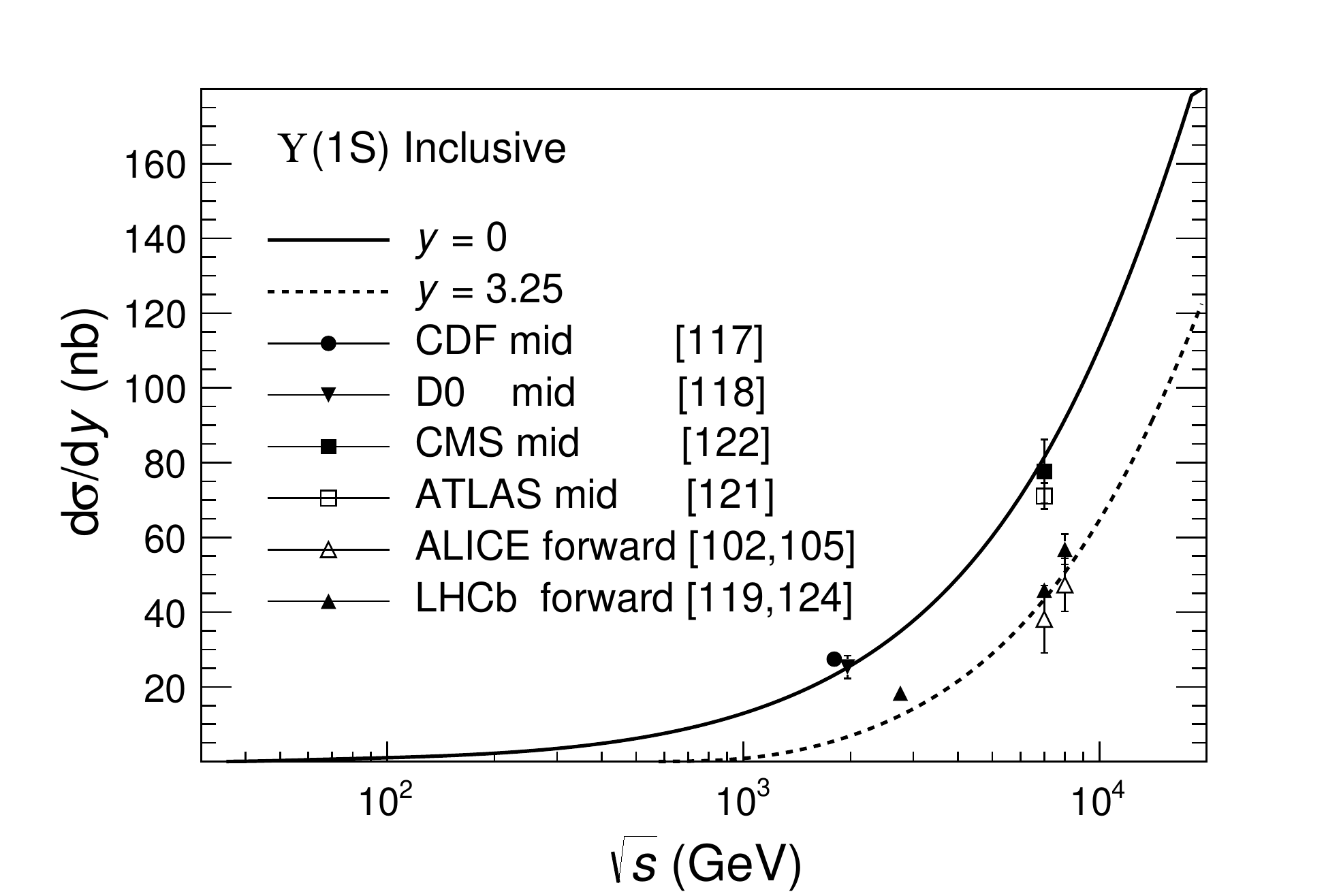}
\end{center}
\vskip -2mm
\caption{$\upsi$ meson $\pt$-integrated cross section as a function of $\en$ at mid-rapidity 
(full line) and forward rapidity of $y = 3.25$ (dashed line) and comparison with the data.}
\label{upsiy0}
\end{figure}

Examples of inclusive $\upsi$ meson $\pt$-spectra fits at different $\en$ values are shown in Fig.~\ref{upsi01} for 
mid-rapidity (left) and forward rapidity of $y = 3.25$ (right). Fig.~\ref{upsiy} presents
the $\upsi$ $\pt$-integrated cross section (for $\pt>0$) as a function of $y$ and comparison with the existing 
measurements at $\en = $ 2.76 and 7~TeV. Our prediction for $\en = 13$~TeV is also given.
The predictions for this cross section dependence on $\en$ at mid-rapidity and at forward rapidity of $y = 3.25$, 
together with the available data, are shown in Fig.~\ref{upsiy0}.

\begin{figure}[ht]
\begin{center}
\includegraphics[width=1.1\columnwidth]{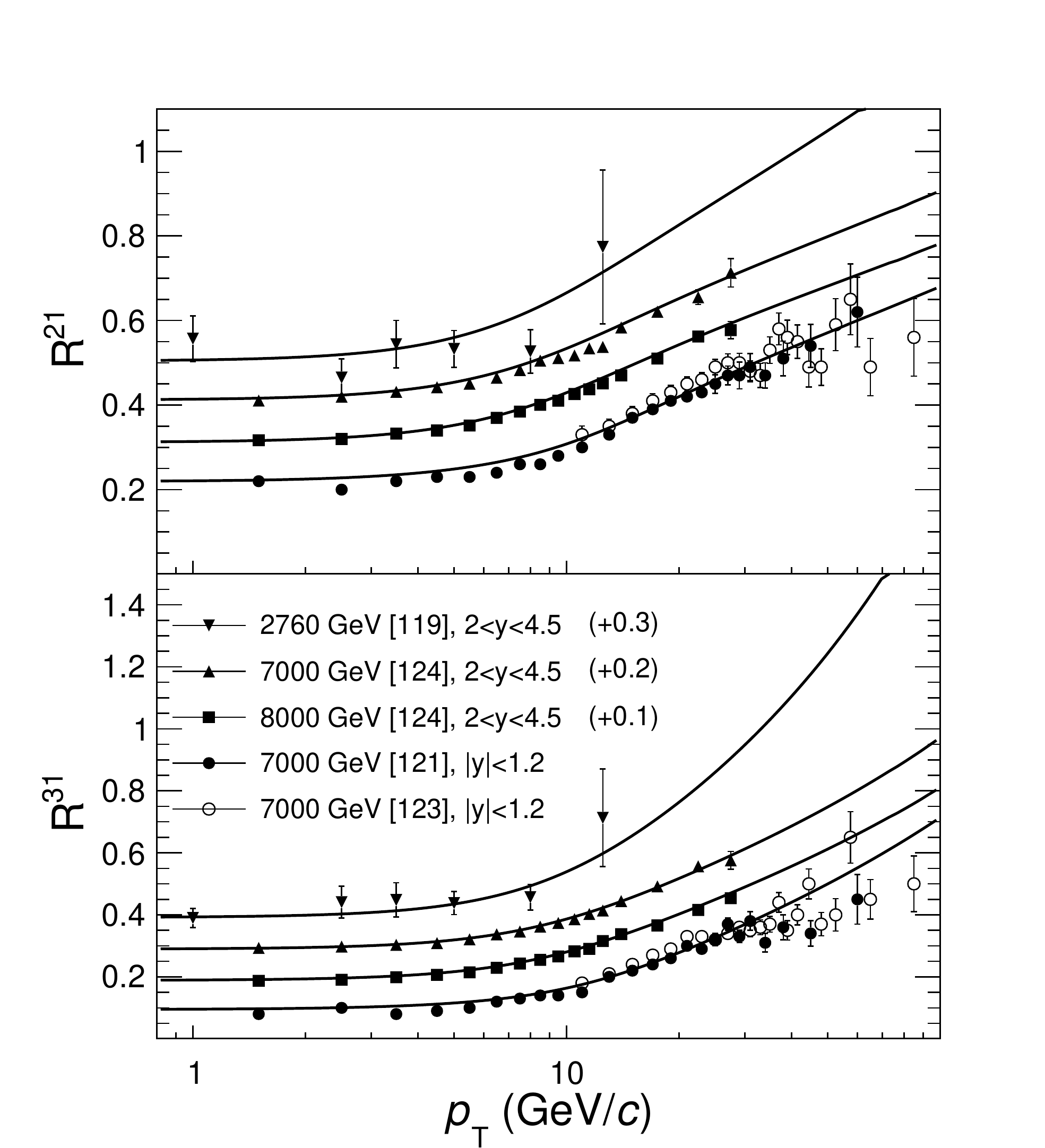}
\end{center}
\vskip -2mm
\caption{Fitting of ratios $R^{21}$ and $R^{31}$ vs $\pt$, described in the text for $\upsip$ and $\upsipp$,
at different $\en$ and $y$ values. Forward rapidity data and lines 
at $\en = $ 8, 7, 2.76~TeV are shifted up by 0.1, 0.2, 0.3, respectively, for a better separation.}
\label{upsiratio}
\end{figure}

\begin{figure}[H]
\begin{center}
\includegraphics[width=1.1\columnwidth]{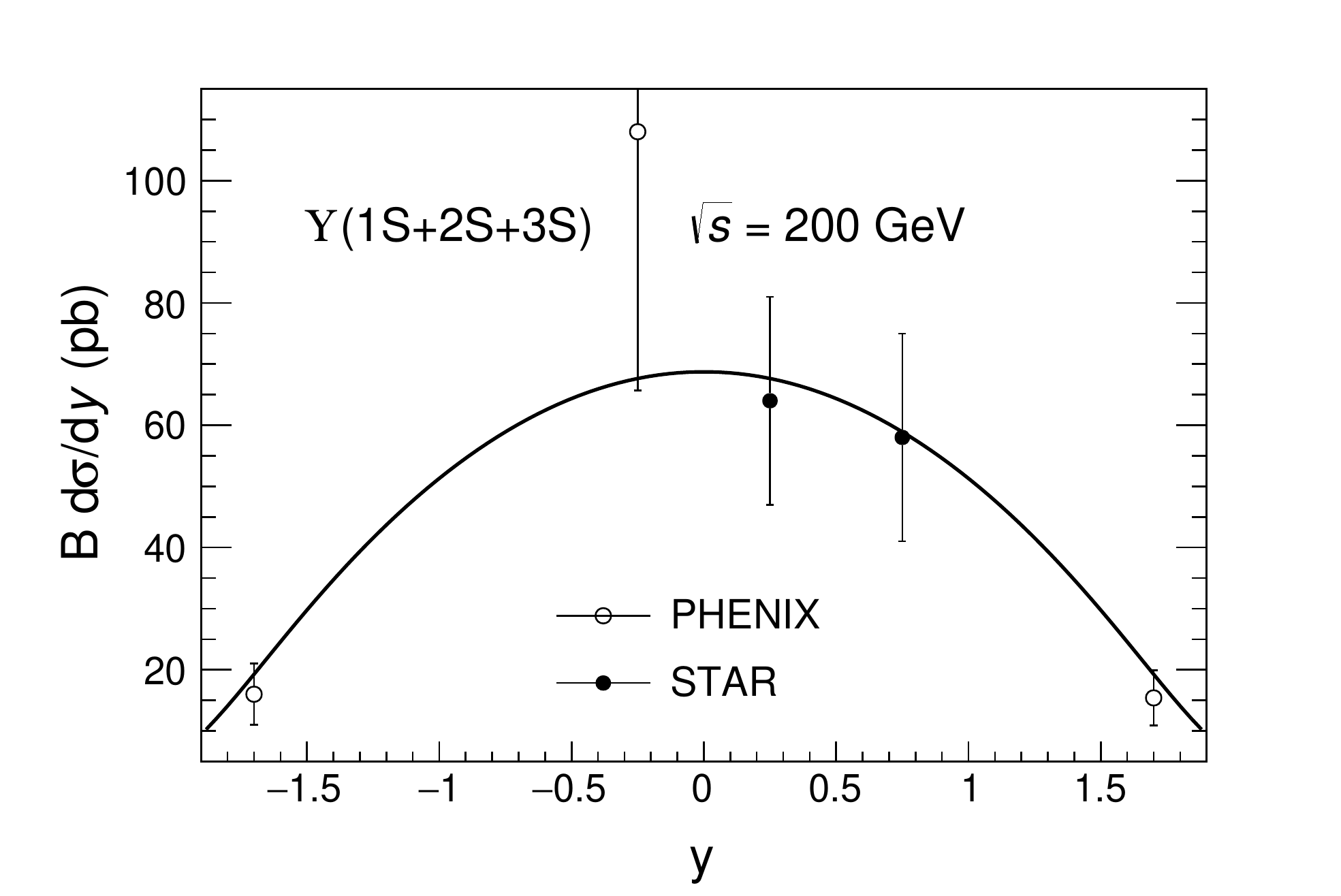}
\end{center}
\vskip -2mm
\caption{$\upsia$ $\pt$-integrated cross section times the dimuon branching fraction $B$ versus $y$ for $\pp$ 
collisions at $\en = 200$~GeV and comparison with the data~\cite{Phen8,Star6,Phen9}.}
\label{upsiy200}
\end{figure}

To illustrate the fits for $\upsip$ and $\upsipp$ mesons, we consider the ratios
of their inclusive production cross sections times their dimuon branching fractions to the same quantity for $\upsi$,
denoted usually as $R^{21}$ and $R^{31}$, respectively.
Figure~\ref{upsiratio} presents the fit results for the $\pt$ dependence of these ratios at different values of $\en$ and $y$.
Lastly, Fig.~\ref{upsiy200} demonstrates the model description of the $\upsia$ production $\pt$-integrated cross section
times the dimuon branching fraction as a function of $y$, measured in $\pp$ collisions at $\en = 200$~GeV~\cite{Phen8,Star6,Phen9}.

\section{\label{sec6}Conclusion}

Thus, we presented a thermal model of a flowing hadronic fireball, based on the TD and BWM, which describes
well almost all available data (except diffractive processes at large $y$ values and some other data sets)
on the pion and quarkonia production yields $d^{2}N/d\pt dy$ in $\pp$ collisions
at $\en \geq 5$~GeV and in $\ppbar$ collisions at $\en > 500$~GeV (where the difference between these two collision 
types can be neglected).
Note that the longitudinal boost invariance is broken in the model due to the used fireball geometry.

One of the distinct features of our model is the assumption that the kinetic freeze-out temperature $T$ is the same 
for all particle types (while their chemical freeze-out temperatures can differ).
$T$ is almost constant at $\en > 500$~GeV, increases with decreasing energy and reaches its maximum at $\en \sim 10$~GeV 
(see Fig.~\ref{T_vsEn}). In this energy region, the parameter $n$ goes to infinity (see Fig.~\ref{n_vsEn}) and 
TD reduces to BGD.
Another feature of the model is that the particle chemical potential $\mu$ is proportional to its mass and vanishes with
increase of $\en$. The nonzero $\mu$ can be interpreted as a measure of the chemical non-equilibrium.
Also, we provide parametrizations for the $\en$ dependence of the model parameters allowing predictions for 
the pion and quarkonia yields in $\pp$ or $\ppbar$ collisions at new energies of the existing and future accelerators. 
An example script is given in~\cite{my1}, showing how one can use our model to compute
these yields at any $\en$, $\pt$ and $y$ in ROOT~\cite{ROOT}.

In our model, the correlation between the parameters $T$ and $q$ (or $n$) and radial flow velocity ($v_s$)
has similar behavior as in other models (see, e.g., \cite{BW1,Tang}). Namely, $T$ and $q$ increase with decreasing $v_s$.
A combined fit of the pion data with $v_s = 0$ gives about 10\% larger $T$ and from 10\% to 70\% larger $q-1$ when
moving from the LHC energies down to $\en \sim 20$~GeV. $\chi^2$ of this fit is about 50\% larger than the one given
in Table~\ref{tab:tab1} for pions. Quarkonia fits also give similar parameter changes. 
It can be seen in Eq.~(\ref{eq:d2N}) that the effect of the radial flow diminishes with increasing rapidity.
Owing to this feature, our model describes the experimental fact (see, e.g., \cite{BW2, LHCb7jpsi, Brahms2}) that the $\pt$ spectrum
of a given particle becomes softer ($\langle \pt \rangle$ becomes smaller) with the increase of its rapidity.

Finally, since the model includes all the ingredients of the thermal source (fireball), it can be applied
for the pion Bose-Einstein correlation studies using, e.g., the methods of~\cite{Bialas14}.

\section*{Acknowledgements}

I thank P.~Dupieux, C.~Hadjidakis, A.~Parvan, O.~Teryaev, M.~Tokarev and G.~Wilk for interest and helpful discussions.
I would also like to thank the anonymous referee for important comments and the suggestion to discuss the correlation between
the model parameters.

\end{document}